\documentclass[11pt]{article}

\usepackage{amsmath}
\usepackage{amsthm}
\usepackage{amssymb}
\usepackage{adjustbox} % rotate table 1
\usepackage{algorithmic}
\usepackage{bbm}
\usepackage{color}
\usepackage{caption,subcaption}
\usepackage{dsfont} % blackboard numerals
\usepackage{graphicx}
\usepackage{hyperref}
\usepackage{mathrsfs}
\usepackage[margin=1in]{geometry}
\usepackage{mathtools}
\usepackage{rotating}
\usepackage{tikz}
\usepackage{wrapfig}
\usepackage{setspace}
\usepackage{booktabs} % To thicken table lines
\usepackage{accents} % to put a circle on top of letters
\usepackage{soul} % strike through texts
\usepackage{relsize} % make a larger summation
\usepackage{lscape} % make certain pages landscape
\usepackage{xcolor} 
\setstcolor{red}
\usepackage{multirow}
\usepackage{booktabs,siunitx}
\newcommand{\indep}{\mathbin{\rotatebox[origin=c]{90}{$\models$}}}

\newcommand{\E}{\mathbb{E}}
\newcommand{\Var}{\text{Var}}

\newcommand{\A}{\mathbf{A}}

\newcommand{\C}{\mathbf{C}}

\newcommand{\Y}{\mathbf{Y}}
\newcommand{\ba}{\mathbf{a}}
\newcommand{\bc}{\mathbf{c}}

\newcommand{\y}{\mathbf{y}}

\DeclareMathAlphabet\mathbfcal{OMS}{cmsy}{b}{n}
\newcommand{\bH}{\mathbfcal{H}}
\newcommand{\bh}{\mathbf{h}}
\newcommand{\bS}{\mathbfcal{S}}

\newcommand{\redsout}{\bgroup\markoverwith{\textcolor{red}{\rule[0.5ex]{2pt}{0.4pt}}}\ULon}

\newtheorem{thm}{Theorem}

\newtheorem{defn}{Definition}
\newtheorem{example}{Example}

\newtheorem{cor}{Corollary}

\newtheorem{assumption}{Assumption}
\newtheoremstyle{break}
  {\topsep}{\topsep}%
  {\itshape}{}%
  {\bfseries}{}%
  {\newline}{}%
\theoremstyle{break}

\usepackage[numbers]{natbib}

\title{Causal estimands and identification of time-varying effects in non-stationary time series from N-of-1 mobile device data}
\author{Xiaoxuan Cai$^{1}$, Li Zeng, Charlotte Fowler$^{2}$, Lisa Dixon$^{3}$, Dost Ongur$^{4}$, Justin T. Baker$^{4}$,\\ Jukka-Pekka Onnela$^{5}$, Linda Valeri$^{2}$\\[1em]
 \normalsize 1. Department of Statistics, The Ohio State University, cai.1083@osu.edu  \\
 \normalsize 2. Mailman School of Public Health, Columbia University\\
 \normalsize 3. Vagelos College of Physicians and Surgeons, Columbia University\\
 \normalsize 4. McLean Hospital \\
 \normalsize 5. T.H. Chan School of Public
Health, Harvard University, onnela@hsph.harvard.edu}
\date{}

\begin{document}
\maketitle

\abstract{Mobile technology (mobile phones and wearable devices) generates continuous data streams encompassing outcomes, exposures and covariates, presented as intensive longitudinal or multivariate time series data. 
The high frequency of measurements enables granular and dynamic evaluation of treatment effect, revealing their persistence and accumulation over time. 
Existing methods predominantly focus on the contemporaneous effect, temporal-average, or population-average effects, assuming stationarity or invariance of treatment effects over time, which are inadequate both conceptually and statistically to capture dynamic treatment effects in personalized mobile health data.
We here propose new causal estimands for multivariate time series in N-of-1 studies.
These estimands summarize how time-varying exposures impact outcomes in both short- and long-term. We propose identifiability assumptions and a g-formula estimator that accounts for exposure-outcome and outcome-covariate feedback. 
The g-formula employs a state space model framework innovatively to accommodate time-varying behavior of treatment effects in non-stationary time series. 
We apply the proposed method to a multi-year smartphone observational study of bipolar patients and estimate the dynamic effect of phone-based communication on mood of patients with bipolar disorder in an N-of-1 setting. 
Our approach reveals substantial heterogeneity in treatment effects over time and across individuals. 
A simulation-based strategy is also proposed for the development of a short-term, dynamic, and personalized treatment recommendation based on patient's past information, in combination with a novel positivity diagnostics plot, validating proper causal inference in time series data.
\\[1em]
}

\section{Introduction}

Mobile devices (smartphones and wearables) have revolutionized the way we collect data and enable real-time collection of extensive individual-specific data, addressing the long-standing barrier of measuring psychological, behavioral, and contextual biomarkers in naturalistic settings for mental health research \cite{heron2010ecological,riley2011health,li2020microrandomized}. 
The use of mobile devices also promotes active engagement and interaction with patients, offering a promising avenue for timely personalized interventions \cite{hayes2014personalized,kholafazad2021smartphone}. 
% Serious Mental Illness (SMI), including bipolar disorder, major depressive disorder and schizophrenia, has emerged as a significant burden of disease, affecting over 13 million individuals in the United States \cite{NSDUH2019}. 
Our research is specifically motivated by the Bipolar Longitudinal Study (BLS) -- a multi-year observational smartphone study of individuals with bipolar spectrum disorder. 

The Bipolar Longitudinal Study (BLS) collected data through smartphones and fitness trackers, including passive information such as GPS traces, accelerometer data, and anonymized summary metrics derived from text message and phone call logs. Participants additionally provided behavioral and psychological self-evaluations through Ecological Momentary Assessment (EMA) \cite{wichers2011momentary,li2020microrandomized}. In contrast to most mobile health studies that span a few weeks or months \cite{li2020microrandomized}, the BLS study follows up patients over multiple years, making it a unique investigation into the long-term effects of behavioral factors and their dynamic evolution. 
Social support has been shown to be crucial in fostering sustained improvements in symptoms and social functioning \cite{johnson1999social,pevalin2003social,corrigan2004social,hendryx2009social}, augmenting antipsychotic treatment to improve patients' quality of life \cite{johnson1999social,lieberman2005effectiveness}. 
We aim to investigate the impact of phone-based communication via text message and phone call logs on negative mood -- a crucial marker for symptom severity. 
Our motivating study stands out as a pioneering endeavor in quantitatively evaluating the dynamic impact of phone-based social connectivity on severe mental illness using mobile technology. %Although the BLS study is unique in its extensive follow-up and rich information collection within the SMI context, we anticipate that long-term close monitoring of patients will become the norm in the future as mobile technology becomes more prevalent, and decision-making tailored to participants' individual health and environmental conditions become an essential ingredient for next generation's health care.
 
The analysis of mobile device data, which involves characteristics of correlated multivariate time series and intensive longitudinal data, introduces specific challenges in study design, non-stationarity, and significant patient heterogeneity. Another persistent challenge in mobile health research lies in the development of appropriate causal estimands that both support personalized decision-making and ensure interpretability. 
This unique context renders a majority of existing designs and causal estimands inappropriate.
Longitudinal studies typically evaluate population-averaged treatment effects of multiple timepoints by assuming a large number of independently and identically distributed (i.i.d) subjects for inference, which is often unsuitable for time series data from mobile devices. 
Traditional time series analysis primarily focus on ``Granger Causality'' -- the association between exposure and outcome at the same time point \citep{granger1969investigating,sims1972money,granger1980testing,granger1988some}, which may suffer from bias in the absence of proper control for confounding  \citep{hsiao1982autoregressive,lechner2010relation,eichler2012causal}. 
\citep{hsiao1982autoregressive,lechner2010relation,eichler2012causal}. 
Various (quasi-) experimental designs for time series (e.g., pre-post test design \cite{imai2019unitfixed}, interrupted time-series design, and difference-in-difference design \cite{imai2021use}) evaluate one-time interventions (also known as ``shocks'') or permanent changes in exposures \cite{frisch1933propagation,slutzky1937summation,ramey2016macroeconomic,bojinov2019time}, making them incompatible with a context of repeated time-varying interventions. 
Cross-sectional time series data (CSTS) has been introduced as a method for tracking i.i.d
subjects over an extended period of time. 
Existing approaches \cite{blackwell2018make,imai2018matching} for CSTS data are tailored to static scenarios, assessing effects averaged across both population and time, and thus are unable to capture dynamic treatment effects in non-stationary contexts.
Sequential multiple assignment randomized trials (SMARTs) target dynamic contemporaneous effect with i.i.d subjects \cite{lizotte2010efficient}, which is insufficient for describing the long-term impacts that are of interest for psychiatric research.
In summary, existing study designs and causal estimands heavily rely on the assumption of large number of i.i.d subjects and a static system for identification, making them ill-suited for mobile health studies characterized by limited subjects and dynamic contexts.
Alternatively, %\citet{rambachan2019econometric} proposed projection-version lagged effects for long-term impact using in non-stationary time series setting with large number of i.i.d subjects.
\citet{bojinov2019time} propose a  ``time series experiment'' for  unit-level temporal average treatment effects with non-parametric identification, however, this approach requires randomization and excludes covariates in consideration.
The N-of-1 design, involving repeated interventions and measurements on a single subject, is a patient-centered study design suitable for observational mobile device data with high patient heterogeneity \citep{kumar2013mobile,daza2018causal}. 
\citet{daza2018causal} proposed  to estimate period-average treatment effects to evaluate the impact of treatments randomly assigned over pre-specified periods, which is not directly applicable in the context of time-varying exposures.
% Observational n-of-1 studies have been largely unexplored due limited availability of data with intensive measurements and long-term follow up, particularly prior to the advancement of mobile health research. 
Novel causal estimands to address the growing demand of individualized inference for dynamic effects in N-of-1 observational studies are urgently needed.

Statistical inference faces additional challenges for intensive mobile device data. First, the densely repeated measurements lead to a higher number of time points than subjects, rendering standard inference based on i.i.d subjects (e.g., generalized linear model), inapplicable \cite{bojinov2019time}. 
While generalized estimation equation (GEE) approaches consider within-subject correlations over time, they have limitations in addressing time series data and do not capture the intricate interdependence structure among the outcome, exposure, and covariates. 
Second, mobile device time series often exhibit non-stationarity due to the changing effect of the treatment and contextual factors. 
Estimating dynamic causal effects for a non-stationary system can be quite challenging \cite{bojinov2019time,stock2016dynamic,ramey2016macroeconomic}. Commonly employed methods for handling non-stationary time series, such as ARIMA and ARCH, present challenges for interpretation due to the complex transformation of the original time series. 
Third, the assignment to exposures in observational mobile health study responds to past outcomes, treatments, and covariates. Temporal feedbacks among treatment, outcomes, covariates is a major obstacle for causal identification with certain variables simultaneously serving as confounders and mediators \cite{hernan2000marginal,imai2018matching,bojinov2019time}. 
Finally, the positivity assumption, a fundamental assumption in causal inference, may be severely violated with large number of exposure time points of interest.
Existing approaches ensure the positivity assumption by randomization \cite{boruvka2018assessing,bojinov2019time,li2020microrandomized}, but there is a lack of tools to evaluate the positivity assumption in observational time series data. 
Thus, for observational mobile device data, it is crucial we develop new estimation strategies that are flexible enough to account for non-stationarity in multivariate time series, high correlation, temporal feedbacks, heterogeneity of participants, and confounding adjustment. 
Furthermore, systematic evaluation tools for the positivity assumption in observational time series data are in need for proper conceptualization of causal estimands.

Our work contributes to the expanding methodological literature on causal inference for intensive longitudinal data and cross-sectional time series data. The objective is to fill a void in causal inference for N-of-1 studies and non-stationary multivariate time series within contemporary personalized mobile device research. We put forward a diverse set of interpretable unit-level causal estimands in N-of-1 studies, elucidating the effects of both single and multiple exposures in both short- and long-term contexts.
% Our approach leverages the potential outcome framework for multivariate time series, allowing us to construct a wide range of interpretable unit-level causal estimands in N-of-1 studies. These estimands cover effects of one exposure and multiple exposures in both short- and long-term. 
Our approach involves the integration of g-formula and mediation analysis techniques to address potential feedback loops among exposure, outcome, and covariates. Additionally, we employ state-space models for personalized statistical inference on both static and dynamic treatment effects, accommodating non-stationarity. 
In addition, we propose two graphical tools -- the ``impulse impact plot'' and ``step response plot'' -- to illustrate the long-term impact of exposure(s) over time. 
Another contribution of our work is a novel positivity assumption diagnostics plot,
which identifies the proper spectrum of treatment regimes that can be identified.
We illustrate the proposed method in the BLS study to evaluate how phone-based social connectivity affects negative mood in the short- and long-term and how these effects change over time. 
Our research provides valuable insights for personalized treatment recommendations in complex mobile health studies. We formulate dynamic and personalized treatment recommendations by leveraging the validation plot for the positivity assumption and considering patients' past treatment and health status. These recommendations are demonstrated through the application to the BLS data. %This article's structure is as follows. Section 2 introduces notations of multivariate time series and directed acyclic graph (DAG) for time series. 

\section{Notation}
We consider an N-of-1 study where a single subject is followed up at discrete times $t=1,2,3,\ldots,T$.
For time $t$, we denote the outcome as $Y_t$, the time-varying exposure(s) or treatment(s) as $A_t$, and other time-varying covariates (e.g., individual and contextual information) as (potential vector) $\C_t$. 
In the Bipolar Longitudinal Study (BLS), $Y_t$ is negative mood at time $t$, $\A_t$ is phone-based social connectivity represented by binary indicators of whether the patient has engaged in communication with at least one close contact via phone calls and text messages at time $t$, and $\C_t$ as a collection of confounders at time $t$ (e.g., physical activity).
The observed series of $\{(A_t,Y_t,\C_t): t \ge 1\}$ constitute a multi-variate time series, with $\C_0$ contains baseline information.
% Without loss of generality, we consider dichotomous exposure $A_t \in \{0,1\}$ and continuous outcome $Y_t$. %We focus on binary exposures; however, out results generalized to categorical or continuous exposures. 
We assume a temporal order defined by treatment $A_t$, outcome $Y_t$, and covariates $\C_t$ within  $t$, such that the administration of the exposure $A_t$ is realized prior to the outcome $Y_t$, which in turn occurs prior to the realization of the covariates $\C_t$. Over the course of the follow-up, the resulting data from an individual are ordered in time as $(\C_0,A_1,Y_1,\C_1,A_2,Y_2,\C_2,\ldots )$, where $\C_0$ contains baseline information. Throughout, we use uppercase letters to represent random variables or vectors and lowercase letters to represent their realized values.

We use subscripts containing a range of consecutive integers to refer to variables or values over the relative subsequent time points. For example, for $q>0$, we denote outcomes from time $t-q$ to $t$ as $\Y_{(t-q):t}=(Y_{t-q},\ldots,Y_t)$ with realized values $\y_{(t-q):t}=(y_{t-q},\ldots,y_t)$; thus, $\Y_{1:t}$ represents the entire outcome path with realized values as $\y_{1:t}$.
Similarly, we denote exposure and covariates from time $t-q$ to $t$ as $\A_{(t-q):t}=(A_{t-q},\ldots,A_t)$ and $\C_{(t-q):t}=(\C_{t-q},\ldots,\C_t)$. 
We denote the data which occurred from $t-q$ to $t$ as history $\bH_{(t-q):t} = \{ (A_s, Y_s,\C_s): t-q \le s \le t \}$, and to simplify, we denote all information up to $t$ as history $\bH_t=\{(A_s, Y_s,\C_s): s \le t \}$. 
In addition, for any variable $X \in \{A_t,Y_t,\C_t: t \ge 0\}$, we denote information occurring prior to variable $X$ as history $\bH_{X^-}$. For example, $\bH_{A_t^-}=\{\A_{1:(t-1)},\Y_{1:(t-1)},\C_{1:(t-1)}\}$ represents the history prior to the administration of exposure $A_t$, $\bH_{Y_t^-} = \{\A_{1:t},\Y_{1:(t-1)},\C_{1:(t-1)}\}$ represents the history prior to the relization of $Y_t$, and $\bH_{\C_t^-}=\{\A_{1:t},\Y_{1:t},\C_{1:(t-1)}\}$ represent the history prior to the realization of $\C_t$.
Adopting notation of causal inference under the counterfactual framework \cite{robins1999estimation,brodersen2015inferring,imai2018matching,blackwell2018make,bojinov2019time,rambachan2019econometric}, we denote $Y_t(a_t)$ as potential outcome under intervention $A_t=a_t$, $Y_{t}(\ba_{(t-q):t})$ as potential outcome under intervention $\A_{(t-q):t}=\ba_{(t-q):t}$, and $Y_{t}(\ba_{1:t})$ as if we were to intervene on the entire exposure path as $\A_{1:t}=\ba_{1:t}$.
% denote the potential outcome path under treatment history $\ba_{(t-q):t}$ as $\Y_{(1:t)}(a_{1:t})= (Y_1(a_1), Y_2(\ba_{1:2}, \ldots, Y_t(\ba_{1:t}))$. 
%can partition the stochastic processes of $\{(Y_t,A_t,\C_t): t \ge 0\}$ into processes occurring prior to $X$ and after $X$ and include the part prior to $X$ in $\bH(X)$.

\begin{figure}
\centering
\includegraphics[width=10cm]{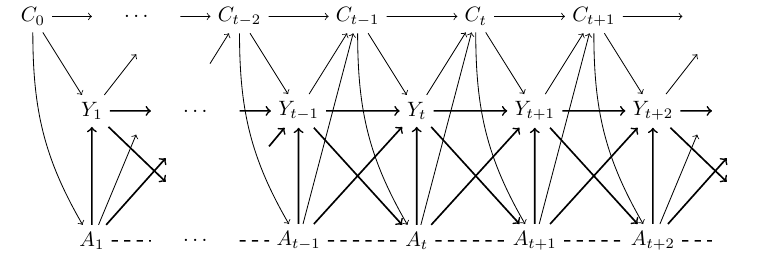}
\caption{Time series directed acyclic graph (DAG) (also known as time series chain graph) displaying the temporal order and relationship of time-varying exposures, outcomes, and covariates. The presence of an arrow indicates a potential causal relationship, while the absence of an arrow indicates there is no causal relationship. Outcome $Y_t$ depends on its previous value $Y_{t-1}$, current and previous exposures $\A_{(t-1):t}$, and previous covariates $\C_{t-1}$; exposure $A_t$ depends on the previous exposure $A_{t-1}$, outcome $Y_{t-1}$, and covariates $\C_{t-1}$; covariate(s) $\C_t$ depend on their previous value(s) $\C_{t-1}$, as well as the most recent exposure $A_{t-1}$ and outcome $Y_{t-1}$. }
\label{fig:dag}	
\end{figure}

We consider a generic relationship for interdependent outcomes, covariates, and exposures over time. At each time $t$, outcome $Y_t$ can be autocorrelated with its previous values and can be affected by current and prior exposures and prior covariates; similarly, exposure(s) $A_t$ can be influenced by prior outcomes, exposures, and covariates, as is common in observational studies \cite{hernan2010causal}; covariate(s) $\C_t$ can be affected by prior covariates, current outcomes and exposures, and influence future outcomes, exposures, and covariates. 
In Figure~\ref{fig:dag}, we present a directed acyclic graph (DAG) (also known as time series chain graph \cite{dahlhaus2003causality}) to show possible causal relationships among the exposure, outcome, and covariates over time for the BLS application, based on previous work \cite{xiaoxuan2022SSMimpute,fowler2022,Valeri2023digitalpsychiatry} and domain knowledge. 
We note that the structure of the DAG can be easily modified to accommodate other research objectives and applications without affecting the identification and estimation of causal estimands below. 
As conventional in causal inference pursuits, researchers should carefully consider the causal structure of their research problem \cite{hernan2023causal}.

Incorporating the entire history can present computational and modeling challenges.
A commonly employed strategy in time series analysis is to assume certain Markov properties, thereby placing constraints on the extent of dependence into the past \cite{imai2018matching,imai2021use}.
We introduce  relevant histories, $\bS_{Y_t^-}$, $\bS_{A_t^-}$, and $\bS_{\C_t^-}$, as subsets selected from $\bH_{Y_t^-}$, $\bH_{A_t^-}$, and $\bH_{\C_t^-}$, respectively. These selected relevant histories contain the necessary information to establish relevant  Markov properties for modeling the distribution of $Y_t$, $A_t$, and $\C_t$.
\begin{assumption}[Markov independence of history] With selected relevant history $\bS_{A_t^-}$, $\bS_{Y_t^-}$ and $\bS_{\C_t^-}$, we have 
\begin{equation}
	\begin{split}
		\text{Pr} (Y_t | \bH_{Y_t^-}) & = \text{Pr} (Y_t | \A_{1:t},Y_{1:(t-1)},\C_{1:(t-1)}) = \text{Pr} (Y_t | \bS_{Y_t^-}) \\
		\text{Pr} (A_t | \bH_{A_t^-}) & = \text{Pr} (A_t | \A_{1:(t-1)},Y_{1:(t-1)},\C_{1:(t-1)}) = \text{Pr} (A_t | \bS_{A_t^-} ) \\
		\text{Pr} (\C_t | \bH_{\C_t^-}) & = \text{Pr} (\C_t | \A_{1:t},Y_{1:t},\C_{1:(t-1)}) = \text{Pr} (\C_t | \bS_{\C_t^-})
	\end{split}
\end{equation}
for all $t>0$.
\label{assp:markov}
\end{assumption} 
Assumption~\ref{assp:markov} states that the distributions of the outcome $Y_t$, exposure $A_t$, and covariates $\C_t$ depend on limited history information, denoted as $\bS_{Y_t^-}$, $\bS_{A_t^-}$, and $\bS_{\C_t^-}$, respectively. %These selected relevant histories are derived from the entire previous history $\bH_{Y_t^-}$, $\bH_{A_t^-}$, and $\bH_{\C_t^-}$ by considering only the subset of information that is relevant for each variable. 
Once we condition on this prior information, variables become conditionally independent with respect to all other prior history. 
As is common in time series analysis, Assumption~\ref{assp:markov}  leads to more efficient and tractable analysis. 
Relevant history is closely tied to the DAG structure. For example, by examining the dependency structure specified by the DAG shown in Figure~\ref{fig:dag}, we have $\bS_{Y_t^-} = \{ \A_{(t-1):t},\C_{t-1},Y_{t-1}\}$, $\bS_{A_t^-}=\{A_{t-1},\C_{t-1},Y_{t-1}\}$, and $\bS_{\C_t^-}=\{ A_t,\C_{t-1},Y_t\}$. 
The choice of past information included in the conditioning set is pivotal, determining how far back in time one must consider confounder adjustment to ensure the validity of sequential exchangeability (Assumption \ref{assp:exchangeability}). Striking a balance is crucial, as an excessive amount of information in the conditioning set may increase model's complexity, while too little information could lead to the omission of vital confounders. In practical terms, we advise researchers to determine the amount of historical information  based on their subject matter knowledge and model selection techniques. It is also prudent to assess the sensitivity of empirical results to the chosen historical information, thereby ensuring a robust and well-informed approach.

\section{Causal Estimands}
\label{section:causalestimands}

Describing the impact of ``an exposure time series'' on ``an outcome time series'' in N-of-1 studies is a challenging task that has remained largely unexplored.
Existing causal estimands, such as population-averaged effects or temporal-averaged effects \cite{imai2018matching,brodersen2015inferring,blackwell2018make}, are unsuitable and inadequate for the dynamic setting of contemporary N-of-1 time series. To illustrate this challenge, consider psychiatric researchers exploring the impact of social connectivity on mood. They may seek to understand: (i) whether social connectivity has an immediate effect on mood, (ii) whether it has a lasting impact over time, (iii) how this impact evolves over time, and (iv) whether these effects interact with other contextual factors, such as physical exercise.

Given this, there is an urgent need to develop appropriate causal estimands that articulate the effects of the exposure on outcomes in both short- and long-term contexts within multivariate time series. These estimands should also elucidate the underlying mechanisms. This work is essential for both N-of-1 studies and intensive longitudinal studies, where traditional estimands fall short in capturing the intricate dynamics of the data.%Historically, time series analysis focuses either on time-invariant contemporaneous effect of exposure on the outcome at the same time point \cite{imai2018matching} or the effect of a one-time intervention (e.g. policy change, advertisement deployment) on the development of outcome \cite{brodersen2015inferring,blackwell2018make};  

%Marginal and conditional causal estimands are essential for understanding the causal relationships between variables over time. 
%Marginal causal estimands present the overall average causal effect of a time-varying exposure on an outcome, regardless of the specific time points or individual characteristics. 
%On the other hand, conditional causal estimands also have gained popularity due to their satisfactory statistical properties \cite{imbens2015causal,bojinov2019time} for  intensive longitudinal studies and time series studies \cite{boruvka2018assessing,imai2018matching,blackwell2018make,rambachan2019econometric,bojinov2019time}. 
%Conditional causal estimands examine the causal effect of a time-varying exposure on an outcome within specific time periods or conditional states, allowing for a deeper understanding of the temporal variations and conditional dependencies present within the time series data. This deeper understanding can inform decision-making, prediction, and intervention strategies in fields such as economics, epidemiology, mental health, and environmental sciences.

Causal estimands regarding recent exposures, rather than entire exposure history, are relevant for N-of-1 studies for several reasons \cite{bojinov2019time}.
First, causal estimands of entire treatment history (e.g., population-average effects in longitudinal studies) are unidentifiable and violate the positivity assumption in N-of-1 studies. 
Second, causal estimands regarding recent exposures align with the ``action in time'' philosophy of N-of-1 studies, as treatment assignments may be dynamically tailored to individual needs and contextual factors.
%Based on pioneering work in cross-sectional time series \cite{imai2018matching,blackwell2018make,rambachan2019econometric} and single unit time series experiments \cite{bojinov2019time}, causal estimands regarding recent exposures, as opposed to  estimands regarding entire exposure history in longitudinal studies, are accepted as valid causal estimands \cite{imbens2015causal} and gaining more attention in modern time series analysis. 
%A typical example is behavioral intervention, as demonstrated by Murphy's group of work, when decisions to send recommendations for physical exercise are highly dependent on the subjects' availability and recent effectiveness of the intervention. 
A general framework for constructing causal estimands regarding recent exposures is to compare the potential outcomes of receiving recent exposures $\ba_{(t-q):t}$ versus $\ba'_{(t-q):t}$ for time points from $(t-q)$ to $t$, as shown below.
\begin{equation}
Y_{t}(\ba_{(t-q):t})-Y_{t}(\ba'_{(t-q):t}) | \bH
\label{eq:generalform_estimand}
\end{equation}
This contrast can be performed while certain historical information in $\bH$ outside the scope of interest is set to its realized values (i.e., conditional causal effects  \cite{imai2018matching,bojinov2019time,boruvka2018assessing}), or $\bH$ is marginalized over (i.e., marginal causal effects \cite{bojinov2019time}).
The duration of treatment of interest, $q$, is not necessarily about determining a correct specification but rather about aligning with a specific research question \cite{imai2018matching}. 
We recommend that researchers select $q+1$ based on their domain knowledge and the objectives of the practical application and carefully consider whether the chosen duration adequately captures the relevant effects and dynamics of interest.

In the following sections, we present a range of causal effects that characterize the effect of repeated time-varying exposures on the development of an outcome over time. We present causal estimands focusing on a single exposure in Section~\ref{sec:effect_one_t} and extend to estimands encompassing multiple exposures over a period of time in Section~\ref{sec:effects_multiple_t}.
We also consider their cumulative effect on all future outcomes in Section~\ref{sec:cumeffects}. 
We note that all causal estimands incorporate a ``time'' component to manifest their dynamic nature.

\subsection{Causal estimands of a single exposure}
\label{sec:effect_one_t}

\begin{figure}[h]
\centering
\begin{subfigure}{.4\textwidth}
    \centering
    \includegraphics[width=\linewidth]{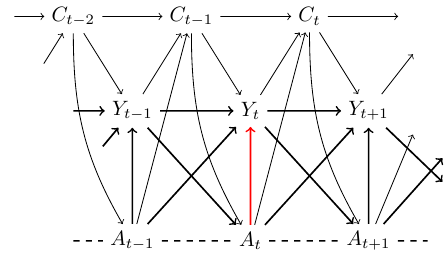}  
    \caption{Contemporaneous effect}
\end{subfigure}
\begin{subfigure}{.4\textwidth}
    \centering
    \includegraphics[width=\linewidth]{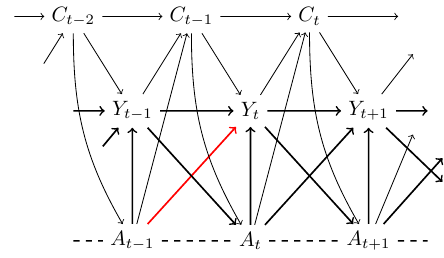}  
    \caption{1-lag structural direct effect}
\end{subfigure}
\vfill
\begin{subfigure}{.4\textwidth}
    \centering
    \includegraphics[width=\linewidth]{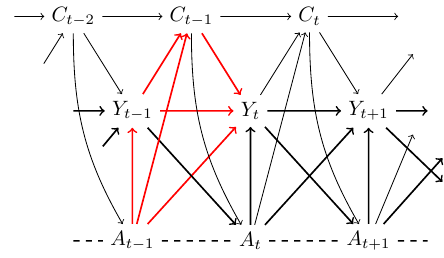}  
    \caption{1-lag effect}
\end{subfigure}
\begin{subfigure}{.4\textwidth}
    \centering
    \includegraphics[width=\linewidth]{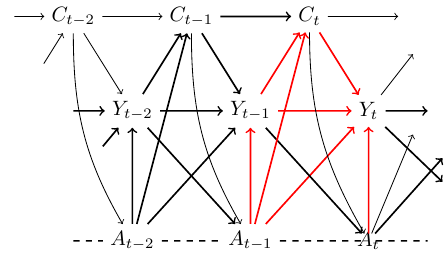}  
    \caption{1-step total effect}
\end{subfigure}
\label{FIGURE LABEL}
\vfill
\begin{subfigure}{.4\textwidth}
    \centering
    \includegraphics[width=\linewidth]{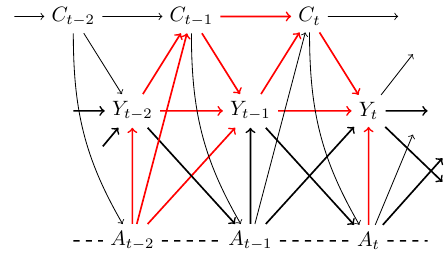}  
    \caption{2-step general effect under $(1,0,1)$}
\end{subfigure}
\begin{subfigure}{.4\textwidth}
    \centering
    \includegraphics[width=\linewidth]{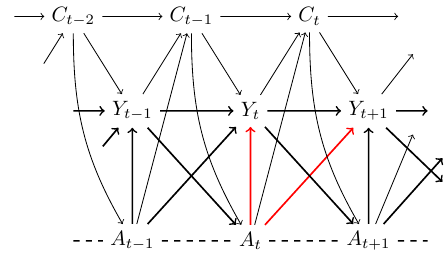}  
    \caption{Cumulative structural direct effect}
\end{subfigure}
\vfill
\begin{subfigure}{.4\textwidth}
    \centering
    \includegraphics[width=\linewidth]{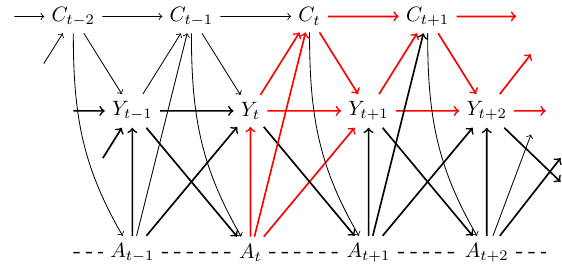}  
    \caption{Cumulative overall effect}
\end{subfigure}
%\begin{subfigure}{.4\textwidth}
%    \centering
%    \includegraphics[width=.95\linewidth]{DAG_cum_general_small.pdf}  
%    \caption{Cumulative general effect under $(1,0,1)$}
%\end{subfigure}
\caption{Graphic representation of causal effects (highlighted by red arrows) in time series directed acyclic graphs (DAGs).}
\label{fig:causal_effects}
\end{figure}

To begin, we consider the causal effect(s) of a single exposure on the outcome, both in the short term (the contemporaneous effect) and in the long term (the q-lag effect). 
\begin{defn}[Contemporaneous effect] The contemporaneous effect of the exposure  $A_t$ on the outcome $Y_t$ is defined as:
\begin{equation}
	\text{CE}_t=Y_t (a_t=1 ) - Y_t ( a_t=0 ).
	\label{def_estimand1}
\end{equation}
\end{defn}
The contemporaneous effect ($\text{CE}_t$), which has received the most attention in time series literature, focuses on the effect of the exposure on the outcome at the same time and describes the difference in the potential outcome $Y_t$ if the individual were treated ($a_t=1$) versus not treated ($a_t=0$). 
For instance, in the context of the BLS study, we can evaluate the effect of phone-based social connectivity on a patient's mood on the same day. 
Figure~\ref{fig:causal_effects}a illustrates $\text{CE}_t$ in a time series DAG.

%\begin{figure}
%\centering
%\includegraphics[width=6cm]{./DAG_DE_small}
%\caption{Contemporaneous effect at time t (in red arrow) in time series DAG.}
%\label{fig:dag_structualDE}	
%\end{figure}
We further explore the direct causal effect of a single exposure on a future outcome, in a scenario where the effect unfolds with a time delay and we do not include any indirect effects through mediating variables. This phenomenon is often observed in situations involving dense measurements or chronic diseases. We define a ``q-lag structural direct effect'' ($\text{LDE}_t^{(q)}$), where $q$ represents the lag duration, capturing the direct causal effect of an exposure at time $t-q$ on a future outcome at time $t$.

\begin{defn}[q-lag structural direct effect]
For $q>0$ and $\bh_{(t-q):(t-1)}\cup \{a_t\} \backslash \{a_{t-q}\}$, the q-lag structural direct effect at time $t$ is defined as:

\begin{equation}
\begin{aligned}
	\text{LDE}_t^{(q)}(\bh_{(t-q):(t-1)}\cup \{a_t\} \backslash \{a_{t-q}\})=Y_t ( a_{t-q}=1, \bh_{(t-q):(t-1)}\cup \{a_t\} \backslash \{a_{t-q}\}) \\ - Y_t ( a_{t-q}=0,\bh_{(t-q):t}\cup \{a_t\} \backslash \{a_{t-q}\}).
\end{aligned}
\label{def_estimand2}
\end{equation}
The set $\{\bh_{(t-q):(t-1)} \cup \{a_t\} \backslash \{a_{t-q}\}\}$ encompasses all variables following the administration of the exposure at $t-q$ and prior to the realization of the outcome $Y_t$ at $t$. By keeping this set fixed, we block all paths from $A_{t-q}$ to $Y_t$ except the direct arrow from $A_{t-q}$ to $Y_t$. 
This allows us to isolate and estimate the direct impact of the exposure $A_{t-q}$ on the outcome $Y_t$, not mediated through any other variable. 
\end{defn}
 This effect is a type of controlled direct effect in mediation analysis. 
In the context of the BLS study, the q-lag structural direct effect can be used to describe the direct impact of  phone-based social connectivity on patient's mood in the future, without considering other intermediate mechanisms.
Figure~\ref{fig:causal_effects}b illustrates the 1-lag structural direct effect $\text{LDE}_t^{(1)}$ in the time series DAG. 
Investigating these q-lag structural direct effects is crucial in constructing an appropriate time series DAG, as it further determines the relevant historical information needed to be included to ensure the sequential exchangeability condition for causal identification. 
In practice, statistical methods, such as sequential testing, causal discovery algorithms and permutation tests \cite{ramdas2019sequential,bojinov2019time}, can also be incorporated to investigate the existence of these structural direct effects and validate the proposed time series DAG.

We additionally examine the overall impact of an exposure on a future outcome, the ``q-lag effect'', considering both the direct impact and the indirect impact through other variables except the exposure at intermediate time points.
\begin{defn}[q-lag effect]
For $q>0$ and $\ba_{(t-q+1):t}$, the q-lag effect at time $t$ is defined as:
\begin{equation}
\text{LE}_t^{(q)}( \ba_{(t-q+1):t})=Y_t (a_{t-q}=1,\ba_{(t-q+1):t})-Y_t (a_{t-q}=0,\ba_{(t-q+1):t}).
\end{equation}
\label{def_estimand3}
\end{defn} 
The q-lag effect ($\text{LE}t^{(q)}$) quantifies the effect of an exposure $A_{t-q}$ on the outcome $Y_t$ if we were to intervene $A_{t-q}$ and compare potential outcomes under treatment ($a_{t-q}=1$) versus control ($a_{t-q}=0$), while holding subsequent exposures constant as $\ba_{(t-q+1):t}$. 
The q-lag effect are closely related to the concept of ``impulse response'' in the time series literature \cite{bojinov2019time, imai2018matching}, which represents the response of the outcome due to a one-time shock or intervention. %, taking into account the subsequent dynamics and interactions within the system.
By estimating the q-lag effect, we can understand the subsequent impact propagating through the system as result of this initial intervention, which is essential for assessing the long-term consequences of exposures in time series data.
In the context of the BLS study, we can use q-lag effect to examine the overall impact of phone-based social connectivity on the patient's mood on a future day.
\begin{example}[1-lag effect] For $a_t=0$ and $q=1$, the 1-lag effect at time $t$ is defined as:
\[\text{LE}_t^{(1)}(a_t=0) = Y_t (a_{t-1}=1,a_t=0) - Y_t (a_{t-1}=0,a_t=0). \]
which captures  the change in the outcome at time $t$ if we were able to change the exposure at time $t-1$ from $a_{t-1}=0$ to $a_{t-1}=1$, while keeping the exposure at the subsequent time point $a_t$ constant at $0$.
The 1-lag effect considers both the structural direct effect of the exposure $A_{t-1}$ on the outcome $Y_t$ ($A_{t-1} \to Y_t$) and indirect effects mediated through other variables ($A_{t-1} \to Y_{t-1} \to Y_t$, $A_{t-1} \to C_t \to Y_t$, and $A_{t-1} \to Y_{t-1} \to C_t \to Y_t$). 
Figure~\ref{fig:causal_effects}c provides a graphical representation of the interventions included in the 1-lag effect $\text{LE}_t^{(1)}(a_t=0)$. In the context of the BLS study, we can use 1-lag effect to examine the overall impact of  phone-based social connectivity on patient's mood the next day.
\end{example}
%\begin{figure}
%\centering
%\includegraphics[width=6cm]{DAG_lag_1_small.pdf}
%\caption{The 1-lag effect at time $t$ (in red arrows) in time series DAG.}
%\label{fig:dag_qlag_1}	
%\end{figure}
%\begin{example}
%Let q=2 and $\ba_{(t-1):t}=(0,0)$, then the 2-lag effect with $\ba_{(t-1):t}=(0,0)$ is defined as,
%\[\text{LE}_t^{(2)}(x_{(t-1):t}=\mathbf{0}) =\E[Y_t(a_{(t-2)}=1,a_{(t-1)}=0,a_t=0)] -\E[Y_t(a_{(t-2)}=0,a_{(t-1)}=0,a_t=0)]\]
%which describes the effect on outcome at $t$ if we were able to flip the treatment at $t-2$ from $a_{t-2}=0$ to $a_{t-2}=1$ while keeping the following exposures $(a_{t-1},a_t)$ constant at $0$. 
%The 2-lag effect above includes indirect influence of $A_{t-2}$ on $Y_t$ carried over via other variables ($Y_{t-2}$, $Y_{t-1}$, $\C_{t-1}$, and $\C_t$) as there is no structural direct effect pointing from $A_{t-2}$ to $Y_t$. Figure~\ref{fig:dag_qlag_2} shows pathways contained in the 2-lag effect $\text{LE}_t^{(2)}$ graphically with red arrows. In the context of BLS application, it can describe the effect of phone-based social connectivity intervention on a patient's mood in the future on day 3, while keeping phone-based social connectivity after the invention back to its normal level.
%\end{example}
%\begin{figure}
%\centering
%\includegraphics[width=10cm]{./Graphs/DAG_lag_2.pdf}
%\caption{Illustration of 2-lag effect ($\text{LE}_t^{(2)}$), represented in red, at time $t$ in time series DAGs.}
%\label{fig:dag_qlag_2}
%\end{figure}

\subsection{Causal estimands of multiple exposures}
\label{sec:effects_multiple_t}
In addition to the causal effects related to an exposure at a single time point, there is a significant interest in the collective influence of causal effects of multiple exposures on the outcome.
This becomes particularly relevant when studying a sequence of exposures or a specific treatment regimen, providing valuable insights on synergies across multiple interventions. 
\begin{defn}[q-step total effect] For $q>0$, let $\mathbf{1}_{q+1}$ and $\mathbf{0}_{q+1}$ denote vectors of 1's and 0's of length $q+1$. The q-step total effect at time $t$ is defined as:
\begin{equation}
\text{TE}_t^{(q)}=Y_t(\ba_{(t-q):t}=\mathbf{1}_{q+1})-Y_t(\ba_{(t-q):t}=\mathbf{0}_{q+1}).
\label{def_estimand4}
\end{equation}	
\end{defn}
The q-step total effect quantifies the change in the potential outcome $Y_t$ at time $t$ resulting from a change in the sequence of exposures from control to treatment that occurred $q$ time points ago. 
This effect takes into account the combined effect of exposures from $t-q$ to $t$. 
In the context of the BLS study, the q-step total effect could be employed to measure the potential impact of a phone-based social connectivity-enhancing intervention program. For instance, we can investigate the effect of consistently engaging in phone-based social connectivity over a course of one week on a patient's mood by the end of the week. 
The $q$-step total effect is closely related to the concept of ``step response'' in the time series literature, which examines the system's response to a permanent change in an exogenous input or shock \cite{brodersen2015inferring,rambachan2019econometric}.
\begin{example}[1-step total effect] The 1-step total effect at time $t$ is defined as:
\[\text{TE}_t^{(1)}=Y_t \left(\ba_{(t-1):t}=(1,1) \right)-Y_t\left(\ba_{(t-1):t}=(0,0) \right),\]
which quantifies the difference in the potential outcome $Y_t$ when receiving treatment compared to control at two time points $t-1$ and $t$. 
Figure~\ref{fig:causal_effects}d provides a graphical representation of the interventions captured in the 1-step total effect $\text{TE}_t^{(1)}$.
\end{example}
%\begin{figure}
%\centering
%\includegraphics[width=6cm]{DAG_qstep_2_small.pdf}
%\caption{The 2-step total effect at time $t$ (in red arrows) in time series DAG.}
%\label{fig:dag_qtotal}	
%\end{figure}

\begin{defn}[q-step general effect] For $q>0$, let  $\mathbf{0}_{q+1}$ denote a vector of 0's  of length $q+1$. For recent exposures $\ba_{(t-q):t}$ from $t-q$ to $t$, the q-step general effect at time $t$ is defined as:
\begin{equation}
\text{GE}_t^{(q)}(\ba_{(t-q):t})=Y_t(\ba_{(t-q):t})-Y_t(\ba_{(t-q):t}=\mathbf{0}_{q+1}).
\end{equation}
\label{def_estimand5}
\end{defn}
The q-step general effect provides a more flexible framework to quantify the relative change in the outcome associated with a customizable sequence of recent exposures, compared to a situation with no interventions at those time points. 
This effect is useful for considering personalized interventions tailored to an individual's need and history.
The q-step general effect extends the concept of the q-step total effect to a more flexible framework of an arbitrary sequence of recent exposures.
\begin{example}[2-step general effect] The 2-step general effect at $t$ under $\ba_{(t-2):t}=(1,0,1)$ is defined as:
\[ \text{GE}_t^{(2)}\left(\ba_{(t-2):t} =(1,0,1)\right)=Y_t \left( \ba_{(t-2):t}=(1,0,1) \right) - Y_t \left( \ba_{(t-2):t}=(0,0,0) \right).\]
which describes the effect on the outcome at $t$ if exposures on recent 3 days were assigned as $\ba_{(t-2):t}=(1,0,1)$, compared to the outcome when exposures were hold constant at zero.
Figure~\ref{fig:causal_effects}e provides a graphical representation of the interventions implied in $\text{GE}_t^{(2)}(\ba_{(t-p):t}=(1,0,1))$.
In the context of the BLS study, we can use general effects to devise an optimal intervention plan. For example, we can compare the 3-step general effects of $(1,0,1)$, $(1,0,0)$, $(0,1,0)$, and $(0,0,1)$ under the constraint of no consecutive intervention in 3 days, and choose the optimal plan by comparing which treatment regimen leads to the greatest improvement in the outcome.
% we can device an optimal intervention plan for the next three days while adhering to the constraint of no consecutive treatments. We can compare the estimated 3-step general effects under different treatment plans, namely $(1,0,1)$, $(1,0,0)$, $(0,1,0)$, and $(0,0,1)$, and determine the optimal intervention plan by evaluating which treatment regimen leads to the most substantial improvement in the outcome on day 3 or over the three-day period on average.
\label{exam:general}
\end{example}
%\begin{figure}
%\centering
%\includegraphics[width=10cm]{./Graphs/DAG_qgeneral_2.pdf}
%\caption{The 2-step general effect under $\ba_{(t-p):t}=(1,0,1)$ at time $t$ (in red arrows) in time series DAG.}
%\label{fig:dag_qgeneral}	
%\end{figure}
\begin{table}
\centering
\begin{tabular}{lll}
  \hline
Causal Estimands & Notation & Explanation \\ \hline
Contemporaneous effect & $\text{CE}_t$ & effect of $A_t$ on $Y_t$ \\
q-lag structural direct effect & $\text{LDE}_t^{(q)}$ & direct effect of $A_{t-q}$ on $Y_t$, not via other variables \\
q-lag effect & $\text{LE}_t^{(q)}(\ba_{(t-q+1):t})$ & total effect of $A_{t-q}$ on $Y_t$ directly and indirectly \\
q-step total effect & $TE_{t}^{(q)}$ & total effect of $\A_{(t-q):t}=\mathbf{1}_{q+1}$ on $Y_t$ \\
q-step general effect & $GE_{t}^{(q)}(\ba_{(t-q):t})$ & total effect of $\A_{(t-q):t}$ on $Y_t$ \\ \hline
\end{tabular}
\caption{Proposed causal estimands for single and multiple exposures. Contemporaneous effect, q-lag structural direct effect, and q-lag effect are used to measure the impact of a single exposure, while q-step total effect and q-step general effect capture the impact of multiple exposures.}
\label{tab:estimands}
\end{table}
Table~\ref{tab:estimands} offers a summary of the proposed causal estimands for both single and multiple exposures, summarizing their definitions, notation, and interpretations.

\subsection{Cumulative causal estimands over time}
\label{sec:cumeffects}
% When considering the impact over all future outcomes over time, we aggregate individual causal effects to construct cumulative effects, which summarizes the long-term influence of the exposure on the trajectory of outcomes.
To assess the impact of the exposure across all outcomes, we aggregate the individual causal effects to establish cumulative effects. This approach measuress the enduring impact of the exposure on the trajectory of future outcomes.
We first consider the cumulative structural direct effect, which summarizes direct effect of the exposure $A_t$ on all future outcomes, while keeping all other intermediate variables fixed.
\begin{defn}[Cumulative structural direct effect] The cumulative structural direct effect for $A_t$ at time $t$ is defined as:
\begin{equation}
\text{cumDE}_t=\text{CE}_t + \sum_{q=1}^{\infty} \text{LDE}_{t+q}^{(q)}.
\end{equation}	
\label{def_estimand6}
\end{defn}
The cumulative structural direct effect accounts for both  the contemporaneous effect ($\text{CE}_t$) and lagged structural direct effects ($\text{LDE}_{t+1}^{(1)}, \text{LDE}_{t+2}^{(2)},\ldots$ for $q=1,2,\ldots$). 
By combining these effects, $\text{cumDE}_t$ captures the total direct change on the entire sequence of outcomes resulting from given an intervention on $A_t$, fixing all other intermediate variables. 
The number of effects included in $\text{cumDE}_t$ depends on the underlying DAG structure of the multivariate time series.

\begin{example}[Cumulative structural direct effect]
For the time series DAG in Figure~\ref{fig:dag}, the cumulative structural direct effect can be calculated as, \[\text{cumDE}_t=\text{CE}_{t}+\text{LDE}_{t+1}^{(1)}\] 
Here, the $\text{cumDE}_t$ consists of two paths: the contemporaneous effect ($A_t \to Y_t$) and the 1-lag structural direct effect ($A_t \to Y_{t+1}$), as shown in Figure~\ref{fig:causal_effects}f. Other lagged structural direct effects $\text{LDE}_{t+q}^{(q)}$ for $q=2,3,\ldots$ are not present and thus are assumed to be $0$. 
In the context of the BLS study, phone-based social connectivity may have a direct impact on today's mood as well as lasting effect on future days' mood through an enhanced sense of support. The cumulative structural direct effect disentangles and quantifies the direct impact of phone-based social connectivity on mood for all subsequent days.
\end{example}
%\begin{figure}
%\centering
%\includegraphics[width=10cm]{./Graphs/DAG_cum_DE}
%\caption{The cumulative structural direct effect at $t$ (in red arrows) in time series DAG.}
%\label{fig:dag_cum_direct}	
%\end{figure}
In addition to the cumulative direct effect, we consider another ``cumulative overall effect,'' which summarizes the overall effect of a single exposure $A_t$ on all future outcomes.
\begin{defn}[Cumulative overall effect] The cumulative overall effect for time $t$ is defined as:
\begin{equation}
\text{cumOE}_t=\text{CE}_t + \sum_{q=1}^{\infty} \text{LE}_{t+q}^{(q)}(\ba_{(t+1):(t+q)}=\mathbf{0} ).
\end{equation}	
\label{def_estimand7}
\end{defn}
%\begin{figure}
%\centering
%\includegraphics[width=10cm]{./Graphs/DAG_cum_qlagged}
%\caption{Components of cumulative overall effect for exposure $A_t$ at $t$ for all future outcomes (in red arrows) in time series DAGs.}
%\label{fig:dag_cum_lagged}
%\end{figure}
The cumulative overall effect combines the contemporaneous effect ($\text{CE}_t$) on outcome $Y_t$ and all lagged effects ($\text{LE}_{t+q}^{(q)}$ for $q=1,2,\ldots$) on future outcomes. 
It sums the impact of an exposure over all future outcomes through all possible pathways that link the exposure to future outcomes.
Figure~\ref{fig:causal_effects}g illustrates all paths involved in the cumulative overall effect in the time series DAGs. 
%In contrast to the cumulative structural direct effect, certain pathways may be counted multiple times with various decay rates for outcomes at various time points. 
%For example, the path of $A_t \to Y_t$ serves as the contemporaneous effect with no decay for the outcome $Y_t$ at time $t$, serves as one segment on the path of $A_t \to Y_t \to Y_{t+1}$ with decay over 1 time point for the outcome $Y_{t+1}$ at $t+1$, serves as one segment on the patch of $A_t \to Y_t \to Y_{t+1} \to Y_{t+2}$ with decay over 2 time points for the outcome $Y_{t+2}$ at time $t+2$, and so forth. Similarly, the path of $A_t \to Y_{t+1}$ serves as the 1-lag structural direct effect with no decay for the outcome $Y_{t+1}$, serves as one segment on the path of $A_t \to Y_{t+1} \to Y_{t+2}$ with decay over 1 time point for the outcome $Y_{t+2}$ at time $t+2$, and so forth. 
% 1-lag effect  ($\text{LE}_{t+1}^{(1)}(a_{(t+1)}=0)$ serves as one component and is contained by all q-lage $\text{LE}_{t+q}^{(q)}(\ba_{(t+1):(t+q)}=\mathbf{0})$ with $q \ge 2$.
% To be more precise, by their definitions, $\text{LE}_{t+1}^{(1)}(a_{(t+1)}=0)$ contains $\text{CE}_t$ as a component in the pathways involved; $\text{LE}_{t+2}^{(2)}(\ba_{(t+1):(t+2)}=\mathbf{0})$ contains all pathways involved in $\text{LE}_{t+1}^{(1)}(a_{(t+1)}=0)$ and $\text{CE}_t$ as components for some of its pathways; and so on for $\text{LE}_{t+q}^{(q)}(\ba_{(t+1):(t+q)}=\mathbf{0})$ with $q \ge 3$. 

Researchers have the flexibility to choose either the cumulative structural direct effect or the cumulative overall effect (or both), depending on the application and scientific interest. For example, in the case of online advertising, the cumulative overall effect is more contextually relevant as it gauges the increase in total sales caused by an advertisement across all future days. 
On the other hand, in the context of the BLS study, as mood fluctuations due to increased social connectivity will eventually diminish over time, the cumulative structural direct effect is more meaningful as it provides valuable clinical insights into the persistence of the intervention's effect. 

\section{Causal Identification}

\subsection{Assumptions}
To identify the potential outcomes defined above, we make the following assumptions for causal identification. While the following assumptions are introduced in the context of N-of-1 studies, many of them are also used for causal inference in intensive longitudinal studies and cross-sectional time series analysis. 

\begin{assumption}[Consistency] For observed recent exposure(s) $\A_{(t-q):t}=\ba_{(t-q):t}$,  \[Y_t = Y_t(\ba_{(t-q):t}) \text{ for } q \ge 0.\] 
\label{assp:consistency}
\end{assumption}
Assumption~\ref{assp:consistency} states % that for recent exposure(s) of interest from $t-p$ to $t$, there is only one version of the exposure, and 
that the potential outcome equals the observed outcome if the recent exposure(s) $\ba_{(t-q):t}$ specified in the potential outcome is the same as the observed exposure(s) $\A_{(t-q):t}$ that the unit received.
\begin{assumption}[Positivity] If the joint density $\Pr(\bH_{t-1}=\bh_{t-1})>0$, then
\begin{small}
	\[
	\Pr(A_t=a_t|\bh_{t-1})>0 \text{ for all } t>0
	\]
\end{small}
for all $a_t$ and $\bh_{t-1}$.
\label{assp:positivity}
\end{assumption}
Assumption~\ref{assp:positivity} states that given any past history, exposure $A_t$ has a change of taking any value at time $t$.
Assumption~\ref{assp:positivity} is satisfied when we allow participants to receive the treatment or control with a positive probability at any time  given any observed history.
\begin{assumption}[Treatment exchangeability] For time $t$ and $q \ge 0$, we have
\[Y_t(a_t) \indep A_t | \bH_{t-1} \]
and 
\[Y_t(\ba_{(t-q):t}) \indep A_k | \bH_{k-1}\] 
for all $k=t-q,\ldots,t$.
\label{assp:exchangeability}
\end{assumption}
Assumption \ref{assp:exchangeability} states that i) given past information of the exposures, outcomes and covariates up to time $t-1$, the treatment $A_t$ is independent of the potential outcome $Y_t(a_t)$; and ii) for $k=t-q,\ldots,t$, given past information of exposures, outcomes and covariates up to time $k-1$, the treatment assignment $A_k$ at time $k$ is independent of the potential outcome $Y_t(\ba_{(t-q):t})$. This assumption will be violated if there are unobserved confounders. 

\begin{assumption}[Intermediate variable exchangeability] For any variable $X$ in the collection of  $X \in \bH_{(t-q):(t-1)}\cup\{\A_t\}\backslash \{A_{t-q}\}$, we have,
\[Y_t(\ba_{(t-q):t}) \indep X | \bH_{X^-}  \]
\label{assp:exchangeability_h}
\end{assumption}
% with multiple mediators. 
This assumption is required for identifying the  q-lag structural direct effect as one type of the controlled direct effect.
Assumption~\ref{assp:exchangeability_h} states that there should be no unmeasured confounding between any of the mediators (including intermediate exposures, outcomes, and covariates in $\bH_{(t-q):(t-1)}\cup\{\A_t\}\backslash \{A_{t-q}\}$) and the outcome $Y_t$, given previous history of each mediator. 
%In the context of time series, confounding variables controlled for mediators ordered in time often overlap, as they all base on slightly different history information; similar assumptions and specification have been commonly used in the structural equation modeling of time series \cite{schultzberg2018number}. 

%the confounding variables for the mediators often overlap, as they share the same history and are influenced by similar factors. 
%Note that the conditioning set in Assumptions~\ref{assp:positivity} and \ref{assp:exchangeability} can be partitioned into three parts: (i) previous exposures before the assignment of $A_k$, $\A_{(t-q):(k-1)}$; (ii) previous time-varying confounders before $A_k$, which include previous covariates $\C_{(t-q):(k-1)}$ and previous outcomes $\Y_{(t-q):(k-1)}$; and baseline information before the entire period of time of interest, $\bH_{k-q}$. 
% This gives Assumptions~\ref{assp:positivity} and \ref{assp:exchangeability} the same format as the conventional positivity and sequential exchangeability used in longitudinal studies.  

Additional assumptions are needed for valid causal inference of N-of-1 studies with a single observed unit \cite{bojinov2019time}.
%of a single-subject time series in N-of-1 studies due to the fact that we only observe one outcome path under one exposure path  \cite{bojinov2019time}. 
To allow for identification with a single subject and dynamic effects, we introduce the following ``periodic stable effects'' assumption. Specifically, we assume that the entire follow-up can be divided into distinct periods, where certain treatment effects remain constant within each period but differ across periods.
Denote $t_{i(j)}$ as the $j$th time point within period $i$ for $j=1,2,\ldots$, so that $t_{i(j)}$ and $t_{i(j')}$ belong to the same period $i$ and $t_{i(j)}$ and $t_{i'(j)}$ belong to different periods for $i \neq i'$. 
\begin{assumption}[Periodic stable effects] 
For $t_{i(j)}$ and $t_{i(j')}$, $j \ne j'$, we have
\[
\text{CE}_{t_{i(j)}}=\text{CE}_{t_{i(j')}} \text{ and } \text{LDE}^{(q)}_{t_{i(j)}}=\text{LDE}^{(q)}_{t_{i(j')}}
\]
for all $i=1,2,\ldots$ and $q>0$.
\label{assp:periodic_stable}
\end{assumption}
Assumption~\ref{assp:periodic_stable} states that the contemporaneous effect and q-lag structural direct effects with $q>0$ are periodic stable and remain the same across all time points within the same period. 
Notably, the partition of these constant effect periods is unknown and must be inferred. 
Assumption~\ref{assp:periodic_stable} neither imposes all effects to be periodic-constant, nor assumes stationarity of the time series.
Rather, it specifically focuses on ensuring the stability of crucial effects (i.e., contemporaneous effect and q-lag structural direct effect) within time intervals.
This approach offers a more realistic and flexible framework for inference, accommodating non-stationarity in dynamic systems while enhancing the interpretability of causal estimands.

\subsection{Causal identification via g-formula}

Causal estimands introduced above in \eqref{def_estimand1} -- (9) involve potential outcomes of the format $Y_t(a_t)$, $Y_t(\ba_{(t-q):t})$, and $Y_t ( a_{t-q}, \bh_{(t-q):(t-1)}\cup \{a_t\} \backslash \{a_{t-q}\})$. We apply the identification assumptions from previous sections to obtain a non-parametric estimator for the causal estimands, also referred to as g-formula \cite{hernan2023causal}.
The g-formula provides a rigorous framework for causal effect identification in observational data, accounting for time-varying confounders, time-dependent treatment assignments, and dynamic causal pathways. 
\begin{thm}[Identification of conditional potential outcomes]
Suppose Assumptions \ref{assp:consistency}-\ref{assp:exchangeability} hold, we have:
\begin{equation}
\E[Y_t(a_t)|\bH_{t-1}] = \E[Y_t|A_t=a_t,\bH_{t-1}] \\
\label{identification_1}
\end{equation}
where $\bH_{t-1}$ is observed history up to time $t-1$. 
Suppose Assumptions \ref{assp:consistency}-\ref{assp:exchangeability} hold and for any $q>1$, we have:
\begin{small}
\begin{equation}
\begin{split}
& \E[Y_t(\ba_{(t-q):t})|\bH_{t-q-1}]  \\
&  = %\sum_{\substack{\y_{(t-q):(t-1)}\\\bc_{(t-q):(t-1)}}} 
\int \E[Y_t|\A_{(t-q):t}=\ba_{(t-q):t},\Y_{(t-q):(t-1)}=\y_{(t-q):(t-1)},\C_{(t-q):(t-1)}=\bc_{(t-q):(t-1)},\bH_{t-q-1}] \\
& \quad \quad \quad \cdot \prod_{k=t-q+1}^{t-1} f(\bc_k|\ba_{(t-q):(k-1)},\y_{(t-q):k},\bc_{(t-q):(k-1)},\bH_{t-q-1}) \\ 
& \quad \quad \quad \cdot \prod_{k=t-q+1}^{t-1} f(y_{k}|\ba_{(t-q):(k-1)},\y_{(t-q):(k-1)},\bc_{(t-q):(k-1)},\bH_{t-q-1}) \\ 
& \quad \quad \quad  \cdot f(\bc_{t-q}|a_{t-q},y_{t-q},\bH_{t-q-1}) \cdot f(y_{t-q}|a_{t-q},\bH_{t-q-1}) \quad  \mathbf{d} \y_{(t-q):(t-1)} \mathbf{d} \bc_{(t-q):(t-1)}
\end{split}
\label{identification_2}
\end{equation}
\end{small}
where $\ba_{(t-q):t}$ are pre-specified exposures, $\y_{(t-q):(t-1)}$ and $\bc_{(t-q):(t-1)}$ are random variables to be marginalized, and $\bH_{t-1}$ is observed history up to time $t-q-1$. For $q=1$, \eqref{identification_2} simplifies to
\begin{equation}
\begin{split}
& \E[Y_t(\ba_{(t-1):t})|\bH_{t-2}]  = %\sum_{\substack{\y_{(t-q):(t-1)}\\\bc_{(t-q):(t-1)}}} 
\int \E[Y_t|\A_{(t-1):t}=\ba_{(t-1):t},Y_{t-1}=y_{t-1},\C_{t-1}=\bc_{t-1},\bH_{t-2}] \\
& \quad \quad \quad \quad \quad \quad \quad \quad \quad \quad \quad  \cdot f(\bc_{t-1}|a_{t-1},y_{t-1},\bH_{t-2}) \cdot f(y_{t-1}|a_{t-1},\bH_{t-2}) \mathbf{d} y_{t-1} \mathbf{d} \bc_{t-1},
\end{split}
\label{identification_2_2}
\end{equation}
Suppose Assumptions \ref{assp:consistency}-\ref{assp:exchangeability_h} hold, for any $q>0$, we have
\begin{equation}
\begin{split}
& \E[Y_t ( a_{t-q}, \bh'_{(t-q):(t-1)}\cup \{a_t\} \backslash \{a_{t-q}\})|\bH_{t-q-1}] = \E[Y_t|A_{t-q}=a_{t-q}, \\
& \quad \quad \quad \quad \quad \quad \quad \quad \bH_{(t-q):(t-1)}\cup \{A_t\} \backslash \{A_{t-q}\} = \bh'_{(t-q):(t-1)}\cup \{a'_t\} \backslash \{a_{t-q}\},\bH_{t-q-1}],
\end{split}
\label{identification_3}
\end{equation}
%\begin{equation}
%\E[Y_t(a_{t-q},\bh_{(t-q):t} \backslash a_{t-q})|\bH_{t-q-1}]= \E[Y_t|A_{t-q}=a_{t-q}, \bH_{(t-q):t}\backslash A_{t-q}=\bh_{(t-q):t}\backslash a_{t-q},\bH_{t-q-1}],
%\label{identification_3}
%\end{equation}
where $a_{t-q}$ is pre-specified exposure, $\bh'_{(t-q):(t-1)}\cup \{a'_t\} \backslash \{a_{t-q}\}$ is pre-specified values of intermediate variables, and $\bH_{t-q-1}$ is observed history up to time $t-q-1$.
\label{thm:identification}
\end{thm}
The detailed proof is shown in the Appendix. To mitigate challenges arising from the computational complexity and model intricacy of incorporating the all history, we leverage Assumption~\ref{assp:markov} and the causal structure to yield a simplified version of Theorem~\ref{thm:identification}.
\begin{cor}[Simplification of conditional potential outcome]
Suppose Assumption~\ref{assp:markov} holds and causal relationship specified as the DAG in Figure~\ref{fig:dag} are true, then we have
\begin{equation}
	\begin{split}
		\E[Y_k|\bH_{Y_k^-}]& = \E[Y_k|\ba_{(k-1):k},\y_{k-1},\bc_{k-1}]\\
		f(\A_k|\bH_{A_k^-})& = f(A_k|\ba_{k-1},\y_{k-1}, \bc_{k-1})\\
		f(\C_k|\bH_{\C_k^-})& = f(\C_k|a_{k},y_{k},\bc_{k-1}).
	\end{split}
\end{equation}
Then supposing Assumptions \ref{assp:consistency}-\ref{assp:exchangeability} hold, the identification of potential outcomes of a single exposure and multiple exposures in \eqref{identification_1}-\eqref{identification_3} can be simplified as: 
\begin{equation}
\E[Y_t(a_t)|\bH_{t-1}] = \E[Y_t|A_t=a_t,A_{t-1},Y_{t-1},C_{t-1}],
\label{eq:simplified_identification_1}
\end{equation}
where $a_t$ is pre-specified, and $A_{t-1},Y_{t-1},C_{t-1}$ are observed values in $\bH_{t-1}$, and for any $q>1$, we have:
\begin{small}
\begin{equation}
\begin{aligned}
& \E[Y_t(\ba_{(t-q):t})|\bH_{t-q-1}]  = \int \E[Y_t|\A_{(t-1):t}=\ba_{(t-1):t},Y_{t-1}=y_{t-1},\C_{t-1}=\bc_{t-1}] \\
& \quad\quad\quad \prod_{k=t-q+1}^{t-1} f(\bc_k|a_{k},y_{k},\bc_{k-1}) f(y_{k}|\ba_{(k-1):k},y_{k-1},\bc_{k-1}) \cdot f(\bc_{t-q}|a_{t-q},y_{t-q},\C_{t-q-1}) \\ 
&  \quad\quad\quad \cdot f(y_{t-q}|a_{t-q},A_{t-q-1},Y_{t-q-1},\C_{t-q-1}) \mathbf{d} \Y_{(t-q):(t-1)}  \mathbf{d} \C_{(t-q):(t-1)} , \\
\end{aligned}
\label{eq:simplified_identification_2}
\end{equation}
\end{small}
where exposures $\ba_{(t-q):t}$ are pre-specified exposures, $Y_{(t-q):(t-1)}$ and $\C_{(t-q):(t-1)}$ are random variables to be marginalized, and $Y_{t-q-1}$ and $\C_{t-q-1}$ are observed values in $\bH_{t-q-1}$, and for $q=1$, \eqref{eq:simplified_identification_2} simplified to
\begin{equation}
\begin{split}
& \E[Y_t(\ba_{(t-1):t})|\bH_{t-2}]  = %\sum_{\substack{\y_{(t-q):(t-1)}\\\bc_{(t-q):(t-1)}}} 
\int \E[Y_t|\A_{(t-1):t}=\ba_{(t-1):t},Y_{t-1}=y_{t-1},\C_{t-1}=\bc_{t-1}] \\
& \quad \quad \quad \quad \quad \quad \quad   \cdot f(\bc_{t-1}|a_{t-1},y_{t-1},\C_{t-2}) \cdot f(y_{t-1}|a_{t-1},A_{t-2}, Y_{t-2}, \C_{t-2})\mathbf{d} y_{t-1} \mathbf{d} \bc_{t-1},
\end{split}
\label{eq:simplified_identification_2_2} 
\end{equation}
and we have: 
\begin{small}
\begin{equation}
\begin{aligned}
& \E[Y_t ( a_{t-q}, \bh_{(t-q):(t-1)}\cup \{a_t\} \backslash \{a_{t-q}\})|\bH_{t-q-1}]=  \\
& \quad \quad \quad \quad \quad \quad \quad \quad  \quad \begin{cases} 
	\E[Y_t|\A_t=a_t, A_{t-1}=a_{t-1}, Y_{t-1}=y_{t-1},C_{t-1}=c_{t-1}] & \textrm{if } q=1 \\
	\E[Y_t|\A_{(t-1):t}=\ba_{(t-1):t}, Y_{t-1}=y_{t-1},C_{t-1}=c_{t-1}] & \textrm{if } q >1 
	\end{cases}
\label{eq:simplified_identification_3}
\end{aligned}
\end{equation}	
\end{small},
where the exposure of interest $a_{t-q}$ and all intermediate variables $\bh_{(t-q):(t-1)}\cup \{a_t\} \backslash \{a_{t-q} \}$ are pre-specified and other information in $\bH_{t-q-1}$ is irrelevant under Assumption~\ref{assp:markov}.
\label{cor:simplification}
\end{cor}
Corollary~\ref{cor:simplification} serves as an example of the simplification derived from Theorem~\ref{thm:identification}, tailored to the DAG structure illustrated in Figure~\ref{fig:dag}. Similar simplifications can be adapted to different DAG structures and relevant history as needed. The detailed proof is shown in the Appendix. 

The potential outcomes identified in Theorem~\ref{thm:identification} are conditional on a particular set of historical information. However, this history may not be of specific interest; rather, we often wish to learn about potential outcomes as a function of exposures alone, by setting the history at its average level, $\E[\bH]$, or marginalizing over an appropriate distribution of history. 
\begin{defn}[Marginalized potential outcomes]
\begin{equation}
	\begin{split}
	\E[Y_t(a_t)] & = \int \E[Y_t|A_t=a_t,\bH_{t-1}=\bh_{t-1}]  \mathrm{d}F_{\bH_{t-1}}(\bh_{t-1}) \\
%	 & = \int_{\s_{A_{t}^-}} \E[Y_t|A_t=a_t,\bS_{A_{t}^-}=\s_{A_{t}^-}]  \mathrm{d}F_{\bS_{A_{t}^-}}(\s_{A_{t}^-}) \\
		\E[Y_t(\ba_{(t-q):t})] & = \int\E[Y_t(\ba_{(t-q):t})|\bH_{t-q-1}=\bh_{t-q-1}] \mathrm{d}F_{\bH_{t-q-1}} (\bh_{t-q-1}) \\
%		& = \int_{\s_{\C_{t-q}^-},\s_{Y_{t-q}^-}} \E[Y_t(\ba_{(t-q):t})|\bS_{\C_{t-q}^-},\bS_{Y_{t-q}^-}] \\
		 \E[Y_t ( a_{t-q}, \bh'_{(t-q):(t-1)}\cup \{a_t\} \backslash \{a_{t-q}\})] & = \int \E[Y_t ( a_{t-q}, \bh'_{(t-q):(t-1)}\cup \{a'_t\} \backslash \{a_{t-q}\})|\bH_{t-q-1}] \\
		 & \quad \quad \quad \quad \quad \mathrm{d}F_{\bH_{t-q-1}} (\bh_{t-q-1}),
	\end{split}
\end{equation}
where $F_{\bH_{t-1}}(\bh_{t-1})$ and $F_{\bH_{t-q-1}}(\bh_{t-q-1})$ are the distribution of the history up to $t-1$ and up to $t-q-1$, respectively. Note that this can either be predefined or estimated from observed data.
\end{defn}

\section{Estimation and inference of dynamic effects using state space model}

Estimating dynamic causal effects in N-of-1 studies introduces unique challenges.
The identification results in Theorem~\ref{thm:identification} and Corollary~\ref{cor:simplification} are entirely non-parametric and can be difficult to estimate in N-of-1 studies when dealing with a large number of time points that exceeds the limited number of units.
 %It can be difficult to identify results in Theorem~\ref{thm:identification} and Corollary~\ref{cor:simplification} in a N-of-1 setting as the number of time points surpasses the number of subjects and large number of variables are considered for modeling. 
To address this issue, we rely on repeated observations from the same individual across time to make individualized inference, assuming certain Markov independence of the past as stated in Assumption~\ref{assp:markov} and periodic stable effects from Assumption~\ref{assp:periodic_stable}.
We propose a state space model framework that can accommodate potential non-stationarity flexibly and incorporate Assumptions~\ref{assp:markov} and \ref{assp:periodic_stable} simultaneously. 
We employ two estimation strategies: parametric modeling, where we fit parametric models  and explicitly quantify causal effects as functions of the estimated parameters; and simulation based, where we impute counterfactuals based on the estimated (parametric or non-parametric) models to measure causal effects. Identification assumptions remain unchanged for both approaches.

% For personalized N-of-1 studies, the effect of exposures on the outcome may be dynamic over time. 
%\citet{} illustrated that the effectiveness of activity-promoting interventions may decrease over time as participants become less receptive to those notifications.
%Therefore, in contrast to the majority of static time series analysis, where stationarity is assumed and causal quantities are assumed to be time-invariant, we allow causal estimands to vary over time and propose to utilize a dynamic linear model (also known as the state space model) for the parameterization of g-formula for potentially non-stationary time series. 
%The simple parameterization under the class of linear models help to simplify the complexity in expressing a broad set of causal effects in the cross-sectional time series setting using a few parameters and their various combinations. 

\subsection{State space model}
State space modeling \cite{aoki2013state} is a well-established technique for estimating time-varying parameters in dynamic systems, facilitated by powerful estimation tools -- Kalman filter and smoothing algorithms \cite{kalman1960new,kalman1961new}. It has found extensive applications across diverse fields, including engineering, economics, statistics, and medicine  \cite{schmidt1966application,harvey1990forecasting,scharf1991statistical,jones1993longitudinal,aoki2013state}. 
The state space model formulates time series $Y_t$, $t=1,2,\ldots$, as observations of a dynamic system's output up to Gaussian random noise. The ``observational equation'' describes the dependence of the observed time series on a latent process of hidden states, while the ``state equation'' describes the evolution of which latent process. 
Denote $\theta_t$ as the $d \times 1$ latent state vector at $t=1,2,\ldots$, and assume it is a Markov process such that  $\theta_t$ is independent of past states $\{\theta_s: s<t\}$, conditional on the previous state $\theta_{t-1}$.

\begin{defn}[Linear state space model] For $t=1,2,\ldots$, the state equation of a linear state space model is defined as
\begin{equation}
\theta_t = G_t \theta_{t-1} + w_t, \quad w_t \sim N_d(0,W_t),
\label{eq:state}	
\end{equation}
where $\theta_{t}$ denotes the $d \times 1$ state vector, $G_t$ is the $d \times d$ state transition matrix, and $w_t$ represents the $d \times 1$ independently and identically distributed noise vector, following distribution  $N_d(0,W_t)$. The observational equation of the state space model is
\begin{equation}
Y_t=F_t \theta_t + v_t,  \quad v_t \sim N_n(0,V_t),
\label{eq:obs}
\end{equation}
where $Y_t$ is the $n \times 1$ vector of  outcomes, $F_t$ is the $n \times d$ observational matrix, and $v_t$ is the independently and identically distributed observational noise vector, following distribution $N_n(0, V_t)$.
\end{defn}
For example, in time series DAGs in Figure \ref{fig:dag}, we consider a one-dimensional outcome $Y_t$ ($n=1$).
The  $F_t$ matrix contains the explanatory variables, including the lagged outcome $Y_{t-1}$, current and previous exposures $(A_t, A_{t-1})$, and current covariates $C_t$. 
Here, the hidden states $\theta_t$ represent the unknown regression coefficients for the explanatory variables at time $t$. 
Then the data generation process of $Y_t$ can be reformulated into a state space model:\begin{equation}
\begin{split}
	Y_t &=  \beta_{0,t} + \rho_t Y_{t-1} + \beta_{1,t} A_t +\beta_{2,t} A_{t-1} + \beta_{c,t} C_{t-1} + v_t	 \\
	& = \begin{pmatrix}
			1 & Y_{t-1} & A_t & A_{t-1} & C_{t-1}
		\end{pmatrix}
		\begin{pmatrix}
			\beta_{0,t} \\ 
		    \rho_t \\
		    \beta_{1,t}\\
		    \beta_{2,t}\\
		    \beta_{c,t}\\
		\end{pmatrix} + v_t
		\\
	&= F_t \theta_t +  v_t,
\end{split}
\label{eq:ssmformat}
\end{equation}
where $F_t =(1,Y_{t-1},A_t,A_{t-1},C_{t-1})'$, $\theta_t=(\beta_{0,t},\rho_t,\beta_{1,t},\beta_{2,t},\beta_{c,t})$, and $v_t \sim N(0,V_t)$. Notably, the coefficients in $\theta_t$ and  variance $V_t$ are functions of time $t$.
A static process occurs if $\theta_t=\dot{\theta}$ is time-invariant; if $A_t$ and $C_{t-1}$ are additional stationary time series, the resulting time series $Y_t$ would also be stationary. 
On the other hand, $Y_t$ may be generated by a dynamic process. 
For example, certain components of $\theta_t$ could be time-varying, such as a random walk, or a periodic stable process, meaning that $\theta_t$ is stationary within periods but varies across periods. We also might consider the variance being time-varying.
The state space model provides a flexible mathematical formulation of both stationary (or static) and non-stationary (or dynamic) time series, and its specification can be easily modified to accommodate complex relationships beyond those illustrated in Figure~\ref{fig:dag}. 
The Kalman filter and smoothing algorithms allow for the efficient estimation of the posterior distribution of coefficients $\theta_t|y_{1:t} \sim N_d(m_t, C_t)$ given observations up to time $t$ and $\theta_t|y_{1:T} \sim N_d(s_t,S_t)$ given all observations, respectively.
\begin{defn}[Kalman Filter] Consider a dynamic linear model specified in \eqref{eq:state} and \eqref{eq:obs} with initial condition $\theta_0 \sim N(m_0,C_0)$. Let 
\[\theta_{t-1}|y_{1:t-1} \sim N_d(m_{t-1}, C_{t-1}).\]
Then the following statements hold:
\begin{equation}
\begin{split}
	m_t &= \E[\theta_t|y_{1:t}] = G_t m_t + R_tF'_t Q_t^{-1}(Y_t - F_tG_tm_{t-1}) \\
	C_t &= \Var [\theta_t|y_{1:t}] = R_t-R_{t}F'_tQ_t^{-1}F_t R_t,
\end{split}	
\label{eq:filter}
\end{equation}
where $R_t=\text{Var}(\theta_{t-1}|y_{1:t})=G_tC_{t-1}G_t'+W_t$ and $Q_t=Var(Y_t|y_{1:t-1})=F_tR_tF'_t+V_t$.
\end{defn}
\begin{defn}[Kalman Smoothing]
Consider the dynamic linear model specified in \eqref{eq:state} and \eqref{eq:obs}. If $\theta_{t+1}|y_{1:T} \sim N_d(s_{t+1}, S_{t+1})$, then $\theta_t|y_{1:T} \sim N_d(s_{t}, S_{t})$, where
\begin{equation*}
\begin{split}
	s_t &= \E[\theta_t|y_{1:T}] = m_t + C_tG'_{t+1} R_{t+1}^{-1}(s_{t+1} - G_{t+1}m_{t}) \\
	S_t &= \Var [\theta_t|y_{1:T}] = C_t-C_{t}G'_{t+1}R_{t+1}^{-1}(R_{t+1}-S_{t+1})R_{t+1}^{-1}G_{t+1}C_t.
\end{split}	
\end{equation*}
\end{defn}
 
\subsection{Estimation using state space model}

State space models serve as powerful tools to address non-stationarity, complex variable relationships over time, and the integration of extensive historical information. 
They enable individualized inference and accommodate crucial modeling assumptions of Markov independence and periodic stability, which underpin robust causal estimation in N-of-1 observational time series studies.
\subsubsection{Parametric regression modeling}
We assume the causal structure outlined in the time series DAGs depicted in Figure~\ref{fig:dag}, translating into relevant histories of $\bS_{Y_t^-} = \{ \A_{(t-1):t},\C_{t-1},Y_{t-1}\}$, $\bS_{A_t^-}=\{A_{t-1},\C_{t-1},Y_{t-1}\}$, and $\bS_{\C_t^-}=\{ A_t,\C_{t-1},Y_t\}$ under Assumption~\ref{assp:markov}. Consequently, we establish the following linear state space models for outcomes and covariates:
\begin{equation}
\E[Y_t|\bH_{Y_t^-}] = \E[Y_t|\A_{(t-1):t},\C_{t-1}, Y_{t_1}] = \beta_{0,t} + \rho_t Y_{t-1} + \beta_{1,t} A_t + \beta_{2,t} A_{t-1} + \beta_{c,t} \C_{t-1}
\label{eq:outcome}
\end{equation}
\begin{equation}
\E[\C_t|\bH_{\C_t^-}]=\E[\C_t|A_t, \C_{t-1}, Y_t] = \mu_{0,t} + \rho_{c,t} \C_{t-1} + \mu_{1,t} A_{t} + \mu_{2,t} Y_{t}.
\label{eq:covariate}
\end{equation}
Note that all parameters $\theta_t=(\beta_{0,t}, \rho_t , \beta_{1,t}, \beta_{2,t}, \beta_{c,t})$ and $\psi_t=(\mu_{0,t},\rho_{c,t},\mu_{1,t},\mu_{2,t})$ are allowed to vary over time, which provides flexibility to accommodate non-stationarity of the system. 
Utilizing the linear parametrization outlined in equations \eqref{eq:outcome} and \eqref{eq:covariate}, we can analytically derive the causal estimands defined above, expressed as functions of the estimated parameters. This approach takes advantage of the statistical properties inherent in the state space model framework, ensuring that these estimations remain asymptotically unbiased. 
%First, we derive causal estimands regarding a single exposure.
\begin{cor}[Estimation of causal estimands regarding a single exposure] Under assumptions~\ref{assp:markov}-\ref{assp:periodic_stable} and the linear state space models \eqref{eq:outcome} and \eqref{eq:covariate}, the contemporaneous effect at $t$ is:
\[
\text{CE}_t =\beta_{1,t},
\]
the 1-lag structural direct effect at time $t$ is:
\[
\text{LDE}_t^{(1)}(y_{t-1},\bc_{t-1},a_t) = \beta_{2,t} \text{ for any } y_{t-1},\bc_{t-1},a_t, 
\]
the q-lag structural direct effect for $q \ge 2$ is:
\[
\text{LDE}_t^{(q)}(\bh_{(t-q):(t-1)} \cup a_t \backslash a_{t-q})  = 0,
\]
the 1-lag effect is:
\[
\text{LE}_t^{(1)}(a_t) = \beta_{2,t}+\beta_{c,t} \mu_{1,t-1} + \rho_t\beta_{1,t-1}+\beta_{c,t} \mu_{2,t-1}\beta_{1,t-1} \text{ for any } a_t,
\]
the 2-lag effect is:
\begin{equation*}
\begin{split}
& \text{LE}_t^{(2)}(a_{(t-1):t}) = (\rho_t + \beta_{c,t}\mu_{2,t-1} )\beta_{2,t-1} + (\beta_{c,t} \rho_{c,t} + \rho_t\beta_{c,t-1}+ \beta_{c,t}\mu_{2,t-1}\beta_{c,t-1})\mu_{1,t-2} \\
& \quad  + (\rho_t + \beta_{c,t}\mu_{2,t-1} )\rho_{t-1}\beta_{1,t-2} +(\beta_{c,t} \rho_{c,t-1} + \rho_t\beta_{c,t-1}+ \beta_{c,t}\mu_{2,t-1}\beta_{c,t-1})\mu_{2,t-2}\beta_{1,t-2},
\end{split}
\end{equation*}
and the 3-lag effect and 4-lag effect as well as their proof are shown in the Appendix.
\label{cor:single_exposure}
\end{cor}

\begin{cor}[Estimation of causal estimands regarding multiple exposures]
Under assumptions~\ref{assp:markov}-\ref{assp:periodic_stable} and the linear state space models \eqref{eq:outcome} and \eqref{eq:covariate}, the 1-step total effect is:
\[\text{TE}_t^{(1)} = \beta_{1,t} + (\beta_{2,t}+\beta_{c,t}\mu_{1,t-1} + \rho_t\beta_{1,t-1}+\beta_{c,t} \mu_{2,t-1}\beta_{1,t-1}) \],
and the 2-step total effect is:
\begin{equation*}
\begin{split}
& \text{TE}_t^{(2)}	= \beta_{1,t} + [\beta_{2,t}+\beta_{c,t}\mu_{1,t}+(\rho_t + \beta_{c,t}\mu_{2,t} )\beta_{1,t-1} ] + (\rho_t + \beta_{c,t}\mu_{2,t} )\beta_{2,t-1} \\
& \quad \quad \quad  + (\beta_{c,t} \rho_{c,t} + \rho_t\beta_{c,t-1}+ \beta_{c,t}\mu_{2,t}\beta_{c,t-1})\mu_{1,t-1} + (\rho_t + \beta_{c,t}\mu_{2,t} )\rho_{t-1}\beta_{1,t-2} \\
& \quad \quad \quad +(\beta_{c,t} \rho_{c,t} + \rho_t\beta_{c,t-1}+ \beta_{c,t}\mu_{2,t}\beta_{c,t-1})\mu_{2,t-1}\beta_{1,t-2},
\end{split}	
\end{equation*}
and the 3-step total effect and 4-step total effect are shown in the Appendix.
\begin{cor}[Estimation of cumulative causal estimand]
Under assumptions~\ref{assp:markov}-\ref{assp:periodic_stable} and the linear state space models \eqref{eq:outcome} and \eqref{eq:covariate}, the cumulative structural direct effect is
\[ \text{cumDE}_t = \beta_{1,t}+ \beta_{2,t}, \]
and the cumulative overall effect is
\[ \text{cumOE}_t = \text{CE}_t+ \sum_{q=1}^{\infty} \text{LE}_{t+q}^{(q)}(\ba_{(t+1):(t+q)}=\mathbf{0}),	
\]
where $\text{CE}_t$ and $\text{LE}_{t+q}^{(q)}(\ba_{(t+1):(t+q)}=\mathbf{0})$ for $q=1,2,\ldots$ are estimated in Corollary \ref{cor:single_exposure}.
\end{cor}
\end{cor}

The identification result above is linked with specific parametric models for outcome and covariates as shown in \eqref{eq:outcome} and \eqref{eq:covariate}. A limitation of this approach is that any modification to the models necessitates a new analytical formulation for estimating causal effects. Alternatively, a Monte Carlo simulation-based method can be considered.

\subsubsection{Monte Carlo simulation}
The Monte Carlo simulation method generates synthetic data from fitted models, offering a flexible approach for assessing causal effects across various model specifications \cite{imbens2015causal}. 
Monte Carlo simulation involves drawing random samples from fitted outcome and exposure models using their respective sampling distributions.
Specifically, for each individual draw, based on the specified exposure(s) in the potential outcomes, we simulate multiple copies of the intermediate variables, following the causal pathways from the exposure(s) to the outcome(s) of interest for assessment, and then simulate the distribution of counterfactual outcomes under the specified exposure(s).
%Using simulated values of intermediate variables and chosen exposures, we further simulate the distribution of counterfactual outcomes under the selected level of exposures.
Causal effects are then estimated by averaging over the simulated counterfactual outcomes of different exposure(s). 
We derive $95\%$ CIs of causal estimands using Monte Carlo sampling of simulated distributions of the counterfactual outcomes.
%The idea is to simulate the distributions of causal estimands by repeatedly simulating potential outcomes using large number of randomly drawn coefficients and other relevant observed history information.  To be more precise, following the estimated state space models for outcome regression $P[Y_t|Y_{t-1},\A_{(t-1):t},\C_t]$ and covariates adjustment $P[\C_t|Y_{t-1},A_{t-1},\C_{t-1}]$, we randomly draw one set of coefficients from their estimated posterior distributions. Based on a particular set of drawn coefficients, we simulate the entire data generation process from time $(t-p)$ to time $t$ with exposures pre-specified as $\ba_{(t-p):t}$, which involves simulating certain intermediate covariates and outcomes. We then obtain one expected $\E[Y_t(x_{(t-p):t})]$ through a series of simulations with the same set of coefficients.
%At last, we repeat the above procedures using different random draws of coefficients to obtain the distribution of $\E[Y_t(x_{(t-p):t})]$. 
To illustrate, we show how to estimate the 1-lag effect in the following example.
\begin{example}[Monte Carlo simulation estimation for 1-lag effect]
Recall that the 1-lag effect at time $t$ is defined as $\text{LE}_t^{(1)}(a_t=0)=Y_t(\ba_{(t-1):t}=(1,0))-Y_t(\ba_{(t-1):t}=(0,0))$. Based on the fitted state space models of outcomes and covariates in \eqref{eq:outcome} and \eqref{eq:covariate}, we simulate the distributions of counterfactuals $Y_t(\ba_{(t-1):t}=(1,0))$ and $Y_t(\ba_{(t-1):t}=(0,0))$ through the steps listed below.
%\[Y_t = \beta_{0,t} + \beta_{1,t} A_t + \beta_{2,t} A_{t-1} + \rho_t Y_{t-1} + \beta_{c,t} \C_{t-1} + \epsilon_t\] 
%\[\C_t = \mu_{0,t} + \rho_{c,t} \C_{t-1} + \mu_{1,t} A_{t} + \mu_{2,t} Y_{t} + w_t \]
We make $K$ random draws of $\Theta_s^{(k)}=\{\beta_{0,s}^{(k)},\rho_s^{(k)},\beta_{1,s}^{(k)}, \beta_{2,s}^{(k)},\beta_{c,s}^{(k)}\}$, and $\Psi_s^{(k)}=\{\mu_{0,s}^{(k)}, \rho_{c,s}^{(k)}, \mu_{1,s}^{(k)}, \mu_{2,s}^{(k)}\}$ for time points $s=t-2,t-1,t$, respectively, from their sampling distributions estimated from the fitted models. 
\begin{itemize}
\item Step 1: For one  draw of $\Theta_s^{(k)}$ and $\Psi_s^{(k)}$, repeat the following steps for copies $b=1,2,\ldots,B$. Based on the observed history of $(A_{t-2},Y_{t-2},\C_{t-2})$ and pre-specified exposure $A_{t-1}=a_{t-1}$, simulate $\tilde{Y}_{(t-1)}^{(k,b)}$ using the fitted model~\eqref{eq:outcome} and $\Theta_{t-1}^{(k)}$. Next, based on the observed $\C_{t-2}$, simulated $\tilde{Y}_{(t-1)}^{(k,b)}$, and pre-specified exposure $A_{t-1}=a_{t-1}$, simulate $\tilde{\C}_{t-1}^{(k,b)}$ using the fitted model~\eqref{eq:covariate} and $\Psi_{t-1}^{(k,b)}$. Finally, based on the simulated $(\tilde{Y}_{t-1}^{(k,b)},\tilde{\C}_{t-1}^{(k,b)})$ and pre-specified $\A_{(t-1):t}=\ba_{(t-1):t}$, simulate the potential outcome $\tilde{Y}_{t-1}^{(k,b)}(\ba_{(t-1):t})$ using the fitted model~\eqref{eq:outcome} and $\Theta_{t}^{(k)}$. 
%Note that $\tilde{Y}_{t-1}^{(k)}$ and $\tilde{\C}_{t-1}^{(k)}$ are intermediate variables to facilitate the simulations of targeted potential outcome. 
\item Step 2: Average over the B copies of simulated $\hat{Y}_{t}^{(k)}(\ba_{(t-1):t})=\frac{1}{B} \sum_{b=1}^B \tilde{Y}_{t}^{(k,b)}(\ba_{(t-1):t})$.
%\item Step 4: Repeat steps 1--3 for  exposure levels $\ba'_{(t-2):t}=(1,0,0)$ and calculate $\hat{Y}_{t}(\ba_{(t-2):t}=(1,0,0))$.
\item Step 3: Apply steps 1-2 for exposure levels of $\ba=(1,0)$ and $\ba=(0,0)$ and calculate the difference of $\hat{Y}_{t}^{(k)}(\ba_{(t-1):t}=(1,0))$ and $\hat{Y}_{t}^{(k)}(\ba_{(t-1):t}=(0,0))$ to be the estimated 1-lag effect.% corresponding to one draw of $\Theta_s^{(k)}$ and $\Psi_s^{(k)}$.
\end{itemize}
Repeat the above procedure for K random draws of parameters $\Theta_s^{(k)}$ and $\Psi_s^{(k)}$. This yields the simulated distributions of % $\hat{Y}_{t}(\ba_{(t-1):t}=(1,0))$ and $\hat{Y}_{t}(\ba_{(t-1):t}=(0,0))$, the difference of which generates the simulated the distribution of 
$\text{LE}_t^{(1)}(a_t=0)$, and a $95\%$ confidence interval for $\text{LE}_t^{(1)}(a_t=0)$ can be computed accordingly. %based on the simulated distribution of $\text{LE}_t^{(1)}(a_t=0)$.
\end{example}
Monte Carlo simulations for other potential outcomes and causal estimands follow similar procedures, and detailed explanations of these simulations are provided in the Appendix. 

%%%%%%%%%%%%%%%%%%%%%%%%%%%%%%%%%%%%%%%%%%%%%%%%%%%%%%%%%%%%%%%%%%%%%%%%%%
%%%%%%%%%%%%%%%%%%%%%% Section 6 : data application %%%%%%%%%%%%%%%%%%%%%%
%%%%%%%%%%%%%%%%%%%%%%%%%%%%%%%%%%%%%%%%%%%%%%%%%%%%%%%%%%%%%%%%%%%%%%%%%%
\section{Application}
\begin{figure}[ht]
\centering
\includegraphics[width=\linewidth]{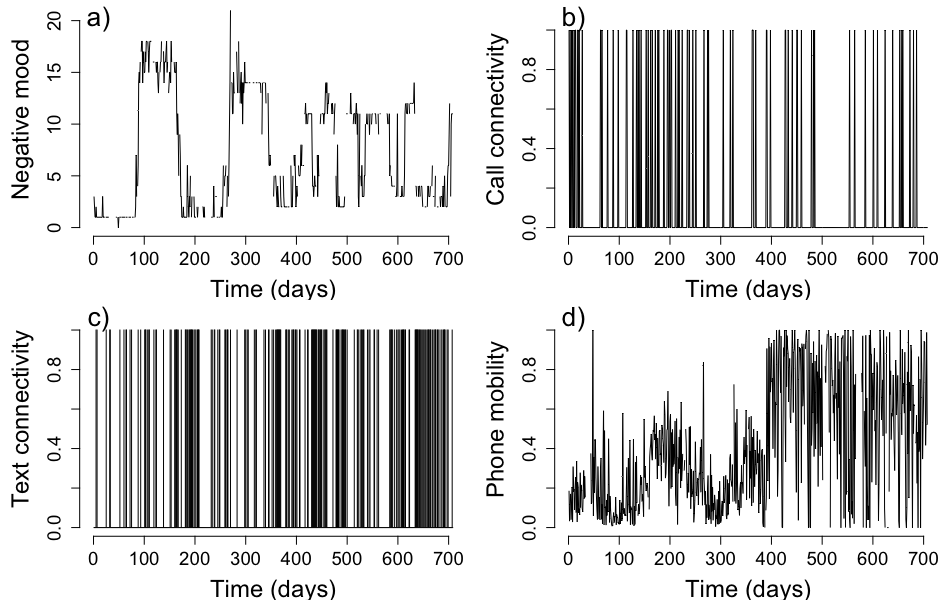}	
\caption{The self-reported negative mood (graph a), binary indicators of phone-based social connectivity with at least one close contact via phone calls (graph b) and text messages (graph c), and phone-derived mobility intensity (graph d) over 708 days of follow up for a bipolar I participant enrolled in BLS.}
\label{fig:raw_data}
\end{figure}

The Bipolar Longitudinal Study is an ongoing mobile health (mHealth) cohort study that has recruited 74 patients with schizophrenia or bipolar disorder from the Psychotic Disorders Division at McLean Hospital since February 2016. 
Once recruited, each subject underwent a comprehensive Diagnostic and Statistical Manual of Mental Disorders (DSM-V) examination. 
A rich collection of data on physical activity, GPS location, and call and text logs was passively collected using smartphones and fitness trackers % without direct involvement from the participants
and a customized 5-minute survey was sent to participants at 5:00 pm every day to inquire about their moods, sleeping, social activities, and psychotic symptoms via the Beiwe platform \cite{huang2019activity,Onnela2021}. 
Here, we examine the effect of phone-based social connectivity on patients' self-reported negative mood, as suggested by past research of the BLS study \cite{xiaoxuan2022SSMimpute,fowler2022,Valeri2023digitalpsychiatry}.  
The outcome of interest -- self-reported negative mood ($Y_t$) -- is a composite measure ranging from 0 (best) to 27 (worst), reflecting a variety of unpleasant emotions including fear, anxiety, embarrassment, hostility, stress, upset, irritation, and loneliness. 
The exposures under investigation are binary indicators of whether the patient has communicated with at least one close contact via phone calls ($A_{calls,t}$) or text messages ($A_{texts,t}$). We identified close contacts using a k-means clustering approach described in the Appendix. Selected contacts satisfy the following criteria: i) mutual communication in both incoming and outgoing directions; ii) frequent communication in both incoming and outgoing directions with the total number exceeding a certain threshold (i.e., 90\% quantile); iii) at least one “long” consecutive text message exchange and one “long” call conversation, defined by exceeding certain thresholds (i.e., 90\% quantile); and iv) communication spanning multiple days \cite{xiaoxuan2022SSMimpute,fowler2022,Valeri2023digitalpsychiatry}. 
Physical activity ($\text{PM}_t$) \cite{peluso2005physical,hamer2012physical} has been demonstrated to be associated with negative mood as well as the tendency for social interaction and is thus included as a confounder. We processed raw phone accelerometer data as a proxy for physical activity, adopting the methodology outlined in \citet{bai2012movelets,bai2014normalization} and denote it as ``phone mobility''.
It is worth noting that covariates, exposures, and outcomes are processed in a way that ensures the collection of exposure precedes the outcome, which in turn precedes the covariates for the same time point $t$, to avoid the potential reverse causation. 
Based on the presumed causal structure in Figure~\ref{fig:dag}, we include current and previous exposures, previous covariates and previous outcome to correct for confounding from auto-correlation in the outcome modeling. 
Similarly, we include current exposure and outcome and previous covariates to correct for auto-correlation in the time-varying covariates modeling. Further discussion considering the inclusion of additional lagged variables is shown in the Appendix.
We describe the assumed data generation process of the outcome and covariate for each time $t$ as follows:
\begin{equation}
\begin{split}
Y_t & = \beta_{0,t} + \rho_t Y_{t-1} + \beta_{11,t} A_{calls,t} + \beta_{12,t} A_{calls,t-1} + \beta_{21,t} A_{text,t} + \beta_{22,t} A_{texts,t-1} \\
	& \quad + \beta_{pa,t} \text{PM}_{t-1} + v_t \\
\text{PM}_t & =  \mu_{0,t} + \rho_{pm,t} \text{PM}_{t-1} + \mu_{1,t} A_{calls,t}  + \mu_{2,t}  A_{text,t} + \mu_{3,t} Y_{t} + u_t.
\end{split}
\label{eq:analysis}
\end{equation}
We demonstrate our proposed method by considering a participant with a Bipolar I disorder diagnosis, and present the detailed analytical results for this participant.
Figure~\ref{fig:raw_data}a shows the observed self-reported negative mood, Figures~\ref{fig:raw_data}b and \ref{fig:raw_data}c show the binary indicators of whether the patient has engaged in communication with close contacts via phone calls and text messages during the day, and Figure~\ref{fig:raw_data}d shows the processed phone mobility data over 708 days of follow up. 
Table \ref{tab:estimation} shows the estimates obtained from the state space models for the outcome and covariate, and a detailed graphical illustrations of the estimated time-varying coefficients across the entire follow-up for all variables are shown in the Appendix.
\begin{table}[ht]
\centering
\begin{tabular}{p{1.7cm}p{1.3cm}p{1.3cm}cp{1.2cm}p{1.3cm}p{1.3cm}c}
  \hline
\multicolumn{4}{c}{Outcome modeling} & \multicolumn{4}{c}{Covariate modeling} \\
\cmidrule(lr){1-4}  \cmidrule(lr){5-8} 
Variable & Estimate & Std.Error &  $90\%$ CI & Variable & Estimate & Std.Error &  $90\%$ CI\\ 
\cmidrule(lr){1-4}  \cmidrule(lr){5-8} 
$\beta_{0,t}$ & \multicolumn{3}{c}{(random walk)} & $\mu_{0,t}$ & \multicolumn{3}{c}{(random walk)} \\ 
$\rho_t$ 				& 0.63*  & 0.04 &   (0.57,0.70)   & $\rho_{\text{pm},t}$(1) & 0.07 & 0.10 & (-0.09,0.23)\\ 
$\beta_{11,t}$			& -0.22 & 0.25 &  (-0.64,0.20) & $\rho_{\text{pm},t}$(2) & 0.18* & 0.09 & (0.03,0.33)\\ 
$\beta_{12,t-1}$ 			& 0.09  & 0.26 &  (-0.34,0.51) & $\rho_{\text{pm},t}$(3)  & -0.11* & 0.05 & (-0.19,-0.04) \\ 
$\beta_{21,t}$(1) & -0.09 & 0.21 &  (-0.43,0.26) & $\mu_{1,t}$ & 0.01 & 0.29 & (-0.47,0.49)\\ 
$\beta_{21,t}$(2) & -1.15* & 0.61 &  (-2.15,-0.15)& $\mu_{2,t}$(1)  & -0.06 & 0.28  & (-0.53,0.41)\\ 
$\beta_{21,t}$(3) & -0.74 & 0.58 &  (-1.70,0.22) & $\mu_{2,t}$(2) & -0.78* & 0.30 & (-1.28,-0.28) \\ 
$\beta_{22,t-1}$(1) & -0.17  & 0.24 &  (-0.56,0.23) & $\mu_{3,t}$ &  -0.01 & 0.03 &  (-0.06,0.03)  \\ 
$\beta_{22,t-1}$(2) & -0.72 & 0.31 &  (-1.24,0.21) \\ 
$\beta_{pm,t}$  		& -0.01 & 0.04 &  (-0.07,0.05) \\ 
   \hline
\end{tabular}
\caption{Estimated coefficients of the outcome state space model (left) and the covariate state space model (right). Statistically significant variables are marked with an asterisk (*). Distinct periods for periodic-stable coefficients are identified by numbers in parentheses. 
Both models identify random walk intercepts ($\beta_{0,t}$ and $\mu_{0,4}$).
In the outcome model, coefficients of $Y_{t-1}$ ($\rho_t$), $A_{\text{calls,t}}$ ($\beta_{11,t}$),  $A_{\text{calls,t-1}}$ ($\beta_{12,t-1}$), and $\text{PM}_t$ ($\beta_{pm,t}$) remain time-invariant. Coefficients for exposures $A_{\text{texts,t}}$ ($\beta_{21,t}$) and $A_{\text{texts,t-1}}$ ($\beta_{22,t-1}$) are periodic stable with change points occurring on day 516 and 641 for $A_{\text{texts,t}}$ and day 461 for $A_{\text{texts,t-1}}$. 
In the covariate model, coefficients for $A_{call,t}$ ($\mu_{1,t}$) and $Y_t$ ($\mu_{3,t}$) remain time-invariant. Coefficients for $\text{PM}_{t-1}$ ($\rho_{pm,t}$) and $A_{text,t}$ ($\mu_{2,t}$) are periodic-stable with change points occurring on day 412 and 470 for $\text{PM}_{t-1}$ and day 412 for $A_{text,t}$.}
\label{tab:estimation}
\end{table}

Note that individualized inferences are conducted for each participant under the framework of an N-of-1 study, given the considerable heterogeneity regarding participants’ enrollment time, length of follow-up, and negative mood trajectory. 
We demonstrate the significant heterogeneity across subjects by comparing four participants in the Appendix. 
The estimated coefficients for exposures (i.e., text and call connectivity with close contacts) and various estimated causal effects differ substantially in magnitude and direction, illustrating the importance of individualized inference to plan pooled analyses properly. 

\subsection{Estimation of causal effects of a single exposure}

\begin{figure}
\begin{subfigure}{0.48\textwidth}
\includegraphics[width=\linewidth]{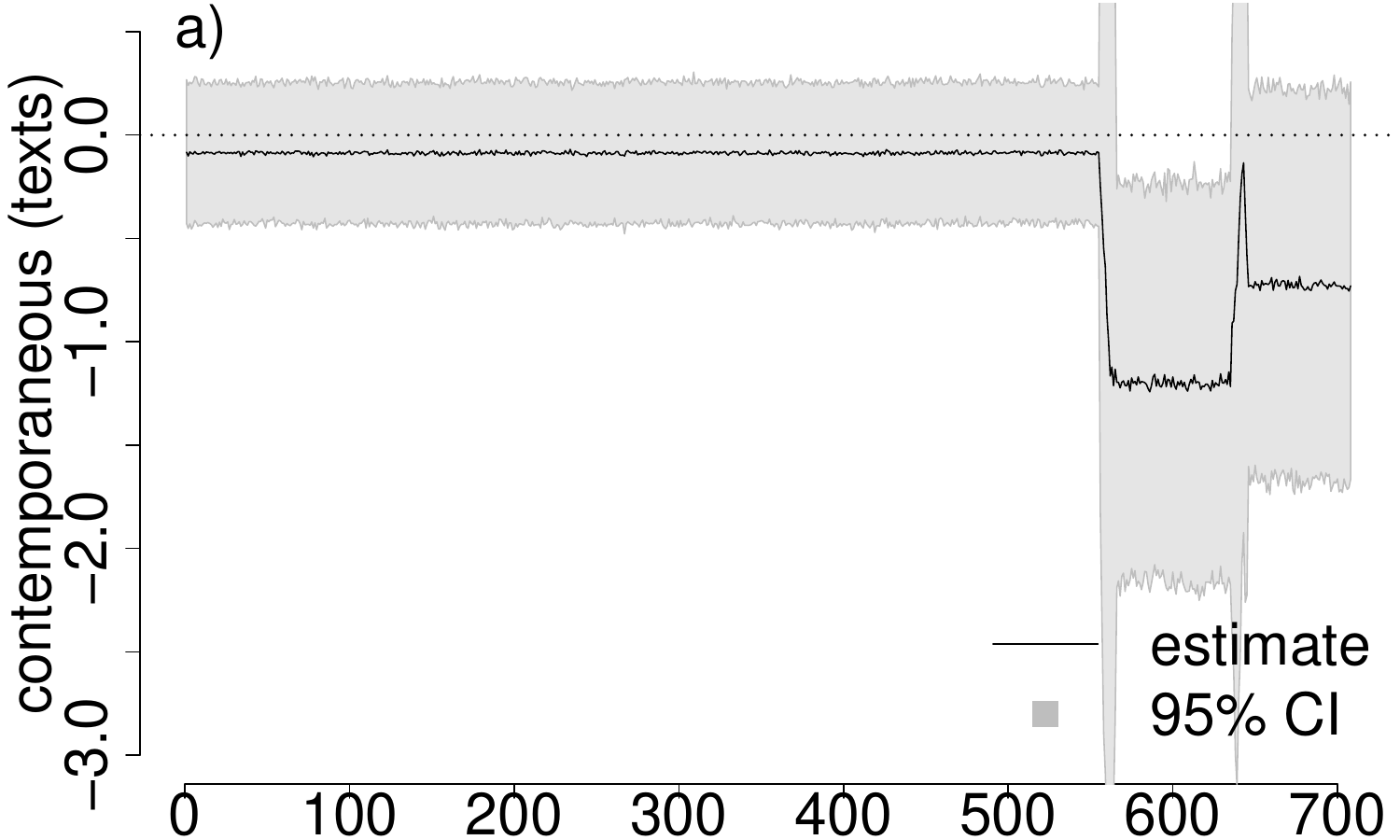}
\end{subfigure}\hspace*{\fill}
\begin{subfigure}{0.48\textwidth}
\includegraphics[width=\linewidth]{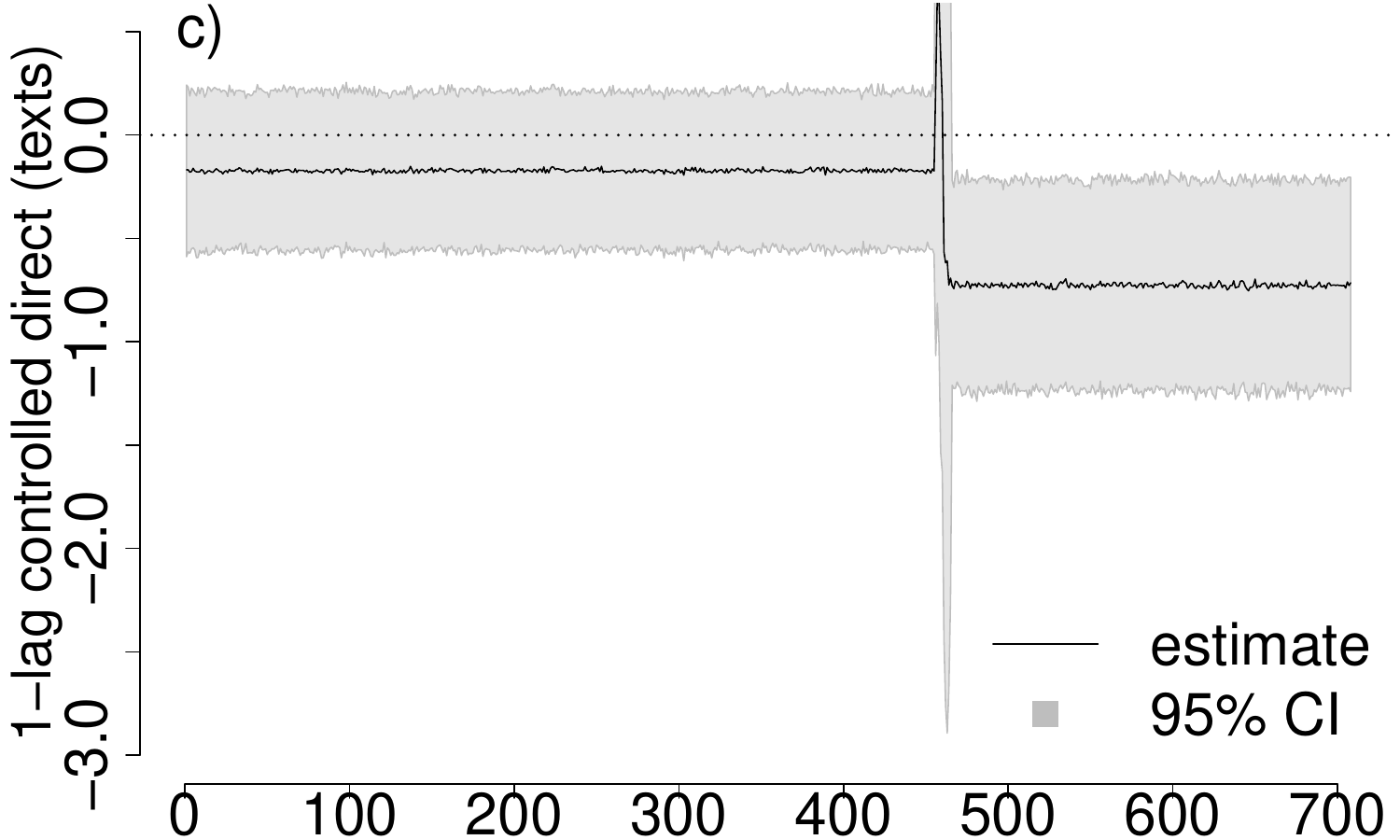}
\end{subfigure}
\medskip %%
\begin{subfigure}{0.48\textwidth}
\includegraphics[width=\linewidth]{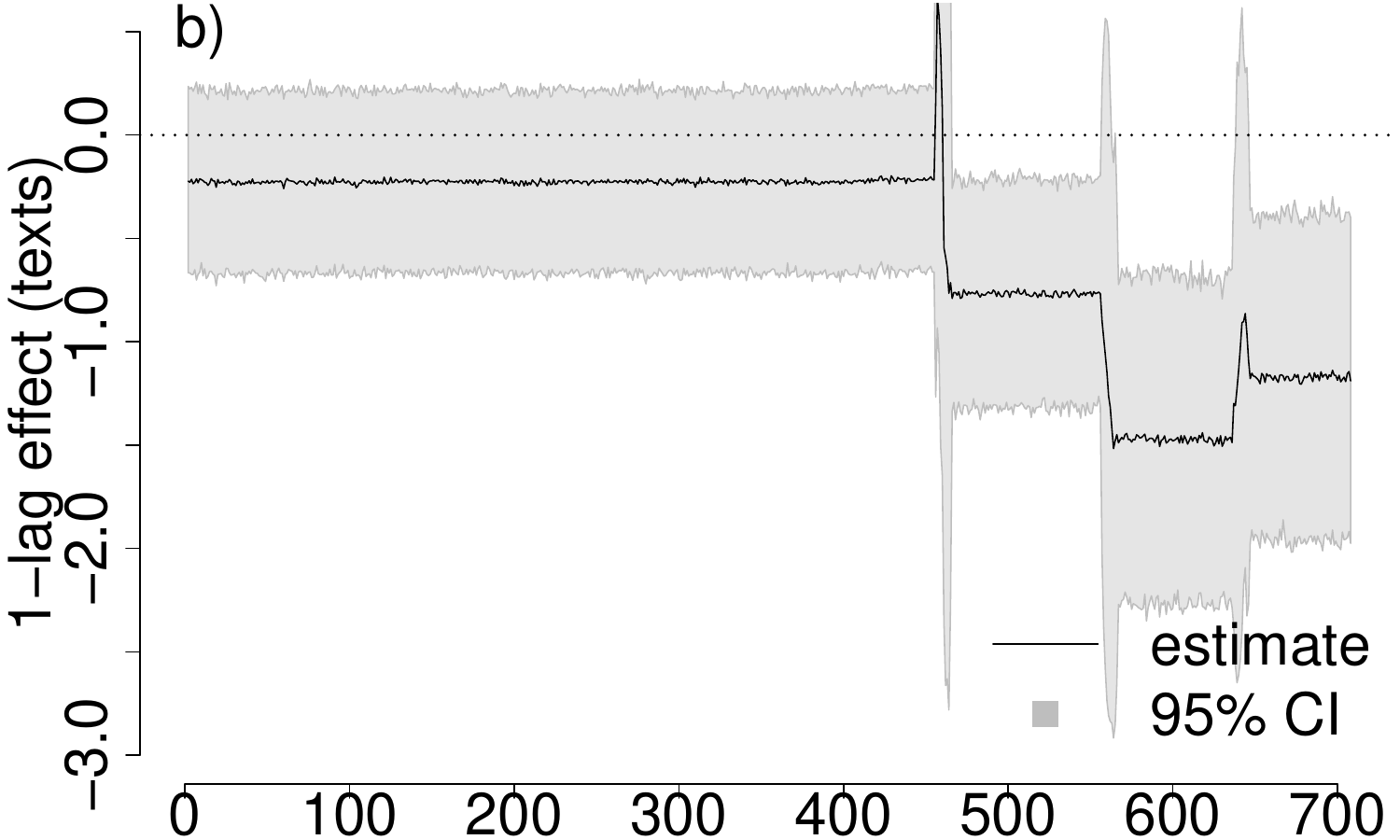}
\end{subfigure}\hspace*{\fill}
\begin{subfigure}{0.48\textwidth}
\includegraphics[width=\linewidth]{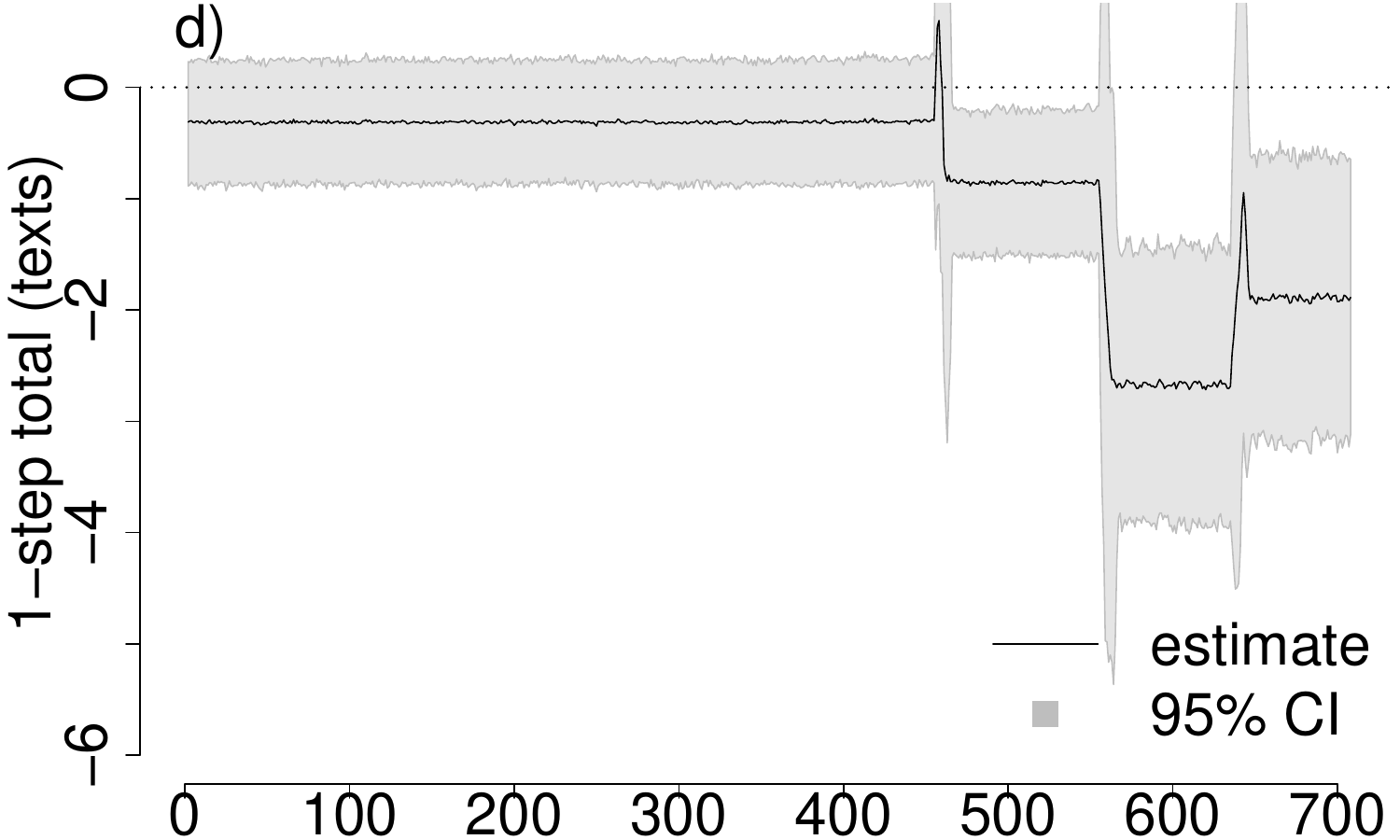}
\end{subfigure}
\caption{Estimated contemporaneous effect (graph a), 1-lag effect (graph b), 1-lag controlled direct effect (graph c), and 1-step total effect (graph d) for text connectivity with at least one close  contact over the follow-up.} 
\label{fig:lag_effects}
\end{figure}

We estimate the proposed causal estimands of the same participant with Bipolar I disorder. 
Figure~\ref{fig:lag_effects} a-c illustrate the estimated contemporaneous effect, 1-lag controlled direct effect, and 1-lag effect for text connectivity over the entire follow-up, all demonstrating a time-varying feature.
% the contemporaneous effect and q-lag effects (for $q=1,2,3$) of digital social connectivity via calls with identified close contacts are all non-significant. This could be due to small sample size that patient is more used to to texting than to making phone calls to her close contacts. 
Specifically, we find that the contemporaneous effect is statistically significant between days 561 and 641, but non-significant prior to or following this period. 
The q-lag effects for $q=1,2,3,4$  are all significant after day 461, with decreasing magnitude as the number of lags increase, corresponding to the diminishing influence of exposure on outcomes over time. 
The 1-lag controlled direct effect is statistically significant after day 461, reconfirming the existence of the arrow from previous exposure $A_{t-1}$ to $Y_t$ in the structure of time series DAGs.
Sensitivity analysis for the q-lag controlled direct effects for $q \ge 2$ are shown in the Appendix, confirming the specification of no additional direct arrows from $A_{t-q}$ with $q \ge 2$ to $Y_t$ in the time series DAGs. 
The 1-step total effect is statistically significant after day 461, combining the contemporaneous effect and 1-lag effect.
Additional estimated causal effects, including i) q-lag effects for $q=2,3,4$ for text connectivity, and ii) contemporaneous effect, q-lag effects, q-lag structural direct effects for $q=1,2,3,4$ for call connectivity over the entire follow-up are shown in the Appendix.

\begin{figure}
\begin{subfigure}{0.48\textwidth}
\includegraphics[width=\linewidth]{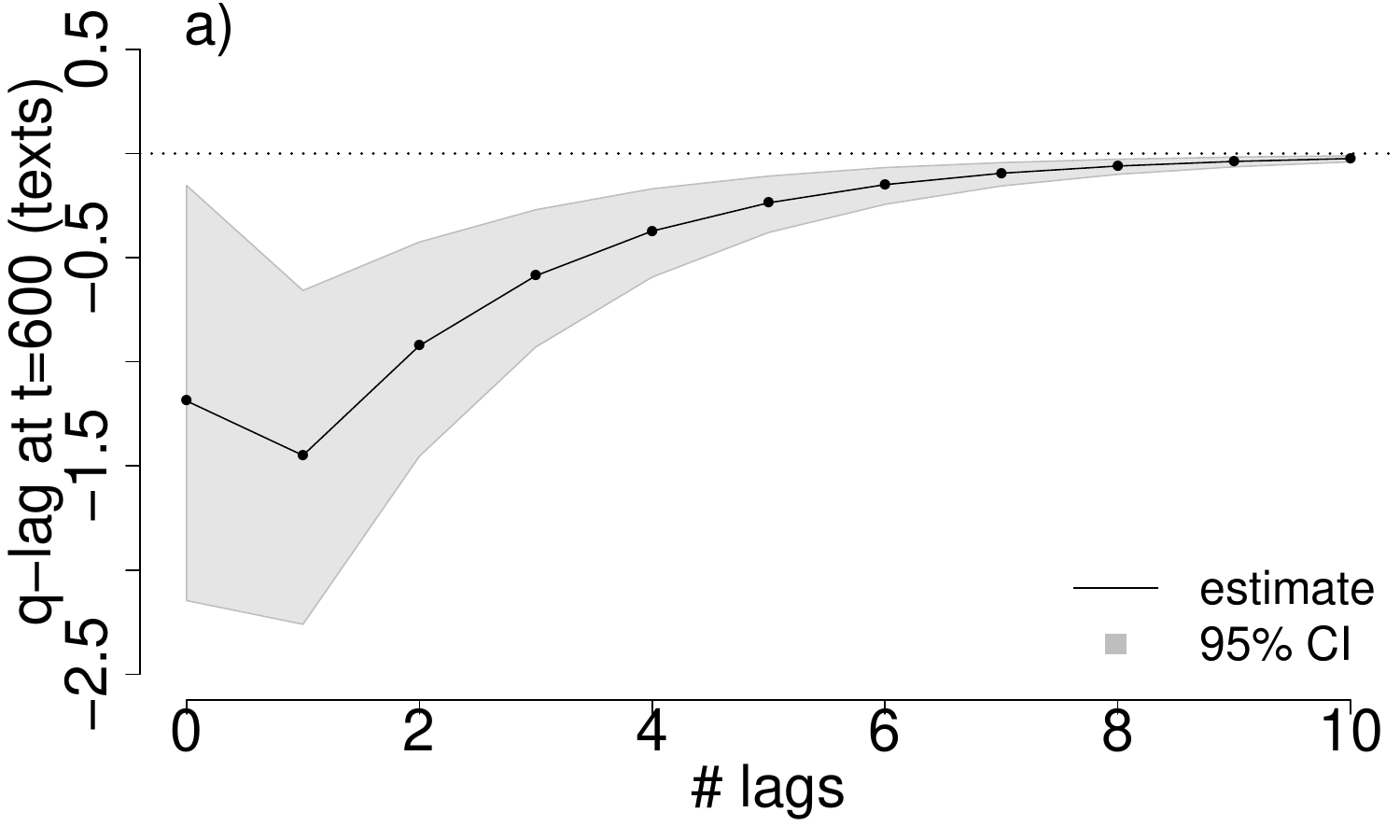}
\end{subfigure}\hspace*{\fill}
\begin{subfigure}{0.48\textwidth}
\includegraphics[width=\linewidth]{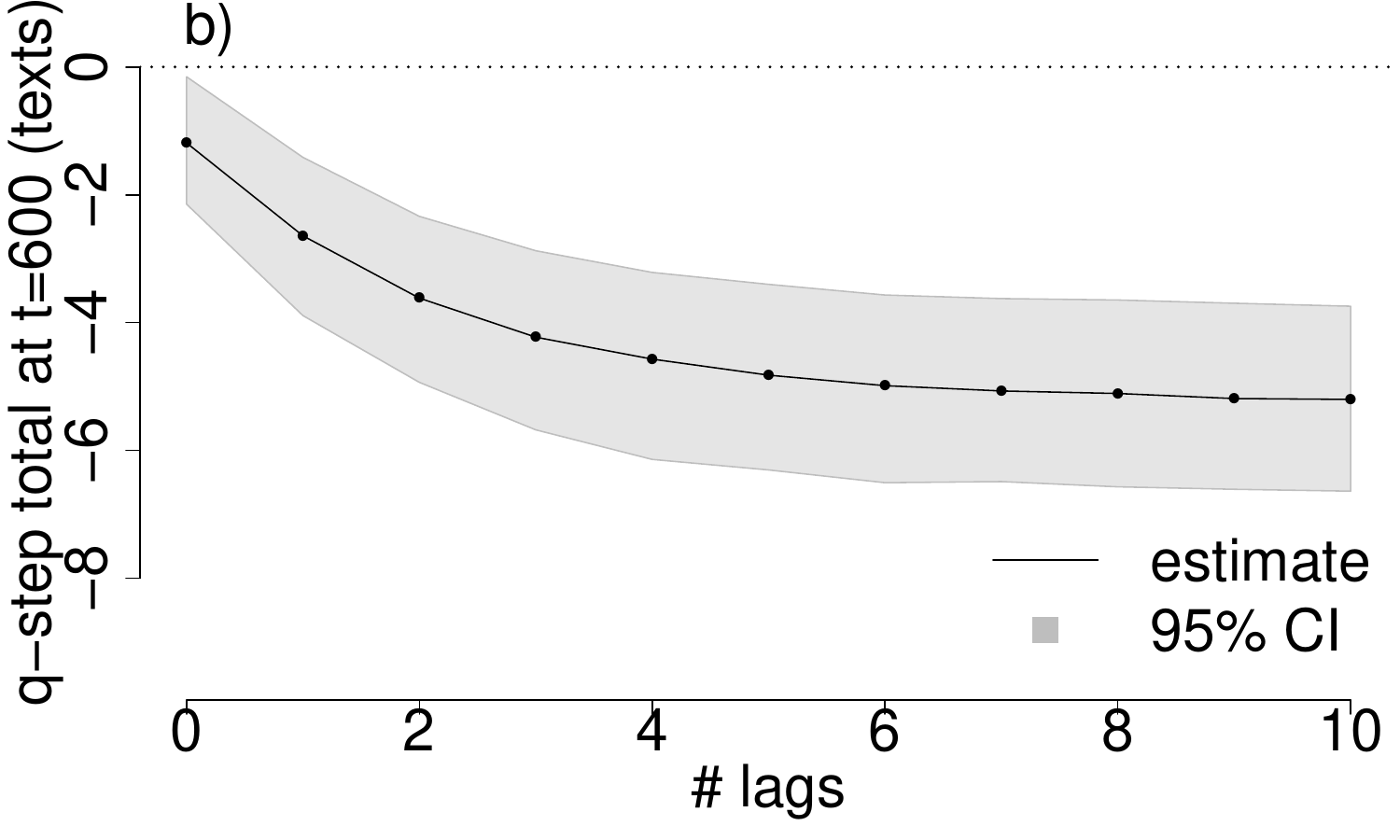}
\end{subfigure}
\caption{Estimated impulse impact plot (graph a) and step response plot (graph b) at day $t=600$ for text connectivity with at least one close contact.} 
\label{fig:two_plots}
\end{figure}

To illustrate the long-term impact of a single exposure on the development of the outcome, we consider an ``impulse impact graph.''
The ``impulse impact graph'' plots the estimated q-lag effects against the number of lags (with $q=0$ corresponding to the contemporaneous effect) and is intended to illustrate the impact of an exposure at $t$ on the development of the outcome and how this impact diminishes over time.
The impulse impact graph in Figure~\ref{fig:two_plots}a depicts the enduring impact of text connectivity at day $t=600$; we observe that its impact initially increases, then decreases, and finally diminishes over time. 

\subsection{Estimation of causal effects of exposures at multiple time points}
To illustrate the effect of a persistent change in the exposure on the development of the outcome, we estimate q-step total effects.
Figure~\ref{fig:lag_effects}d illustrates the estimated 1-step total effect for text connectivity over the follow-up.
Additional estimated q-step total effects for text connectivity for $q=2,3,4$ and call connectivity for $q=1,2,3,4$ over the entire follow-up are shown in the Appendix.
We additionally employ the ``step response graph'' to depict the long-term impact of a permanent change in the exposure.
The ``step response graph'' plots q-step total effects against the number of lags (with $q=0$ corresponding to the contemporaneous effect) and is intended to illustrate the impact of a permanent change in the exposure from placebo to treatment on the development of outcome over time. 
Figure~\ref{fig:two_plots}b illustrates the ``step response function'' for text connectivity at $t=600$ for up to $q=10$ lags. 
We observe that the impact of text connectivity gradually accumulates over time, achieving $80\%$ of its maximum effect after 4 days and $95\%$ after 7 days. This observation suggests a potentially noteworthy intervention window of 4 days for future intervention design.

\begin{figure}[ht]
\begin{subfigure}{0.48\textwidth}
\includegraphics[width=\linewidth]{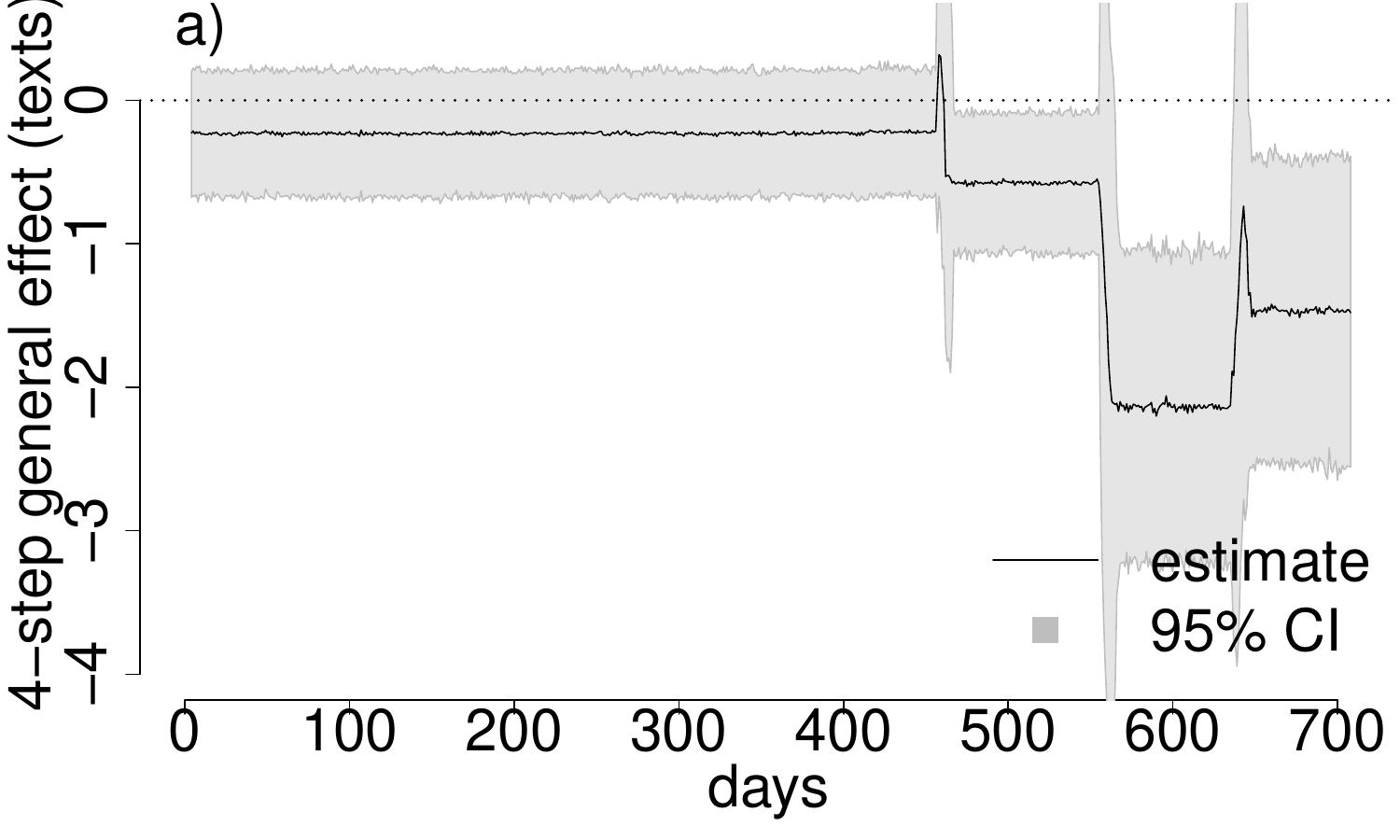}
\end{subfigure}\hspace*{\fill}
\begin{subfigure}{0.48\textwidth}
\includegraphics[width=\linewidth]{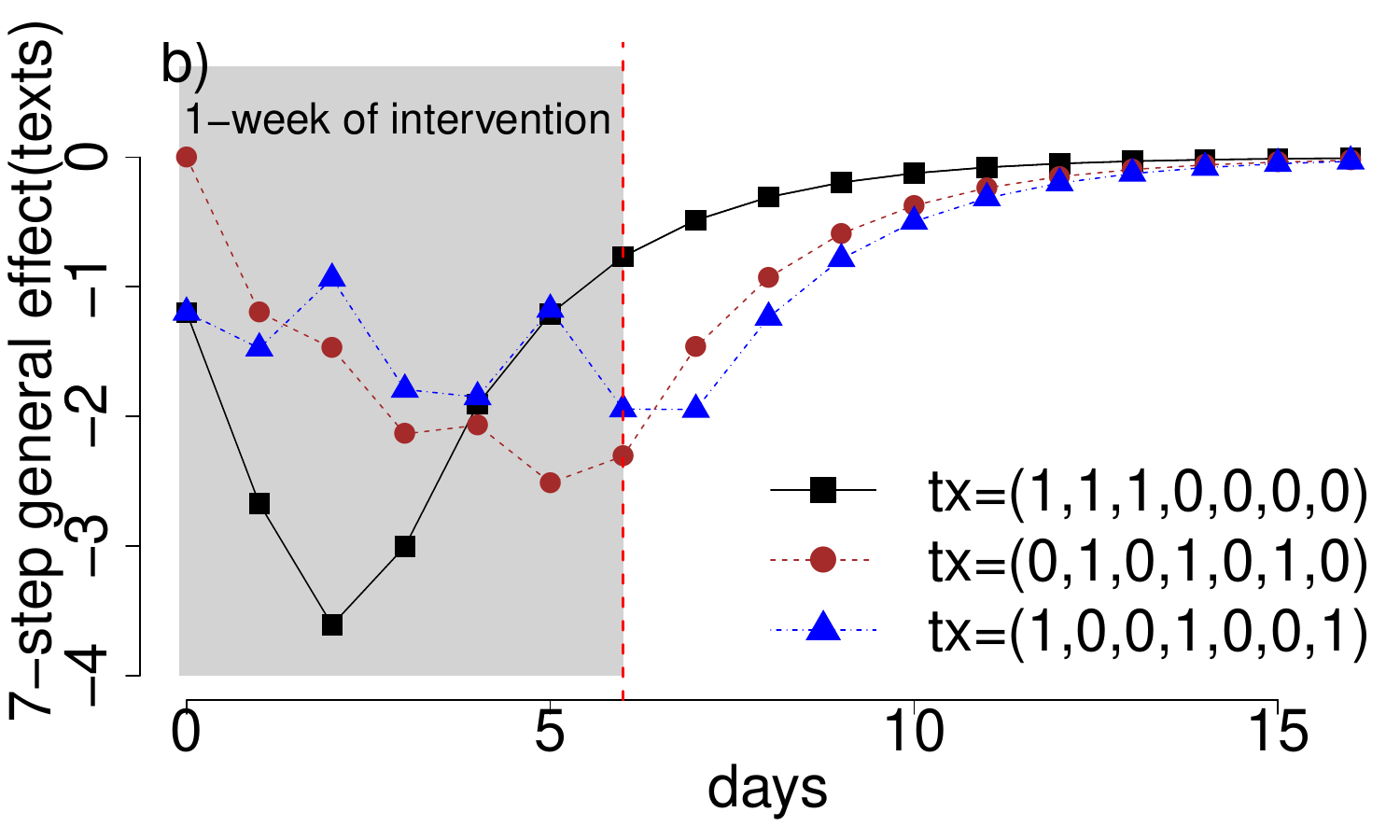}
\end{subfigure}
\caption{Estimated 4-step general effect of a four-day intervention $\text{GE}_{600}^{4}(0,1,0,1)$ at $t=600$ (graph a) and the general response plot for three distinct 7-day interventions starting at $t=600$ under the constraint of no more than three interventions per week (graph b).} 
\label{fig:generaleffects}
\end{figure}

For a general intervention strategy, we estimate q-step general effects and employ the ``general response plot'' to depict the long-term impact. 
Figure~\ref{fig:generaleffects}a illustrates estimated 4-step general effect of a four-day intervention $\text{GE}_t^{(q)}(\ba_{(t-3):t})=(0,1,0,1)$, where text connectivity happens on days 2 and 4 but not on days 1 and 3. 
The 4-step general effect $\text{GE}_t^{(q)}(\ba_{(t-3):t})=(0,1,0,1)$ is statistically significant after day 461. 
Furthermore, in Figure~\ref{fig:generaleffects}b, the general response plot demonstrates three distinct strategies for a 7-day intervention, under the constraint that only three interventions per week. 
This constraint is representative of real-world limitations, including financial and effort constraints.
For example, a psychiatrist may choose to engage with a patient three times in the week following psychotherapy. 
Figure~\ref{fig:generaleffects}b) illustrates the impact of the three intervention strategies.
We find that strategy $\ba_{(t-6):t}=(1,1,1,0,0,0,0)$, or deploying an early intervention on the first three days, yields the quickest improvement in outcomes; however, this improvement diminishes rapidly after the intervention, resulting in the smallest long-term effect after the seven-day period.
Alternatively, strategy $\ba_{(t-6):t}=(0,1,0,1,0,1,0)$, or assigning interventions every other day, leads to a gradual improvement less dramatic and also more lasting effect after the seven-day period.
Finally, strategy $\ba_{(t-6):t}=(1,0,0,1,0,0,1)$, or assigning interventions every other two days, provides the most steady improvement over the course of 7 days and the largest long-term effect among the three strategies considered.
%The above strategies are only examples of strategies that satisfy a hypothetical constraints. 
It is important to note that we refrain from providing a specific recommendation for an optimal strategy. If the objective is to rapidly alleviate symptoms and prevent severe events, strategies with the fastest short-term improvement may be preferred. Conversely, if the objective is to achieve stable improvement with minimal mood fluctuation, strategies with less dramatic improvement may be more suitable. 
We recommend that the researcher carefully consider the ethical and practical constraints and seek out the optimal personalized strategy that aligns with the goals of their investigation.

\subsection{Positivity assumption for various length of exposures}
The assumption of positivity is integral to ensure the validity of causal inferences, however its integrity might be significantly compromised, particularly when considering the effect of a large number of exposure time points of interest. 
To facilitate the proper construction of causal estimands with reliable causal identification, we introduce the ``positivity validation plot.'' 
This plot serves to assess the duration and values of recent exposures which would satisfy the assumption of positivity. Specifically, consider $q$ binary exposures from $t-q+1$ to $t$, denoted as $\A_{(t-p+1):t}$. We calculate the number of observed distinct interventions over all possible $2^p$ values of the intervention. 
If all $2^p$ values are observed, the positivity percentage is $100\%$, ensuring the positivity assumption for any recent exposures of length $p$ in causal inference. 
However, if certain values are not observed, the positivity percentage is less than $100\%$, rendering causal identification potentially invalid for exposure sequences with no observed values due to a potential structural violation of positivity. We recommend researchers to select treatment strategies within an exposure duration with an associated valid positivity assumption that also best suits their needs.

\begin{figure}[ht]
\begin{subfigure}{0.48\textwidth}
\includegraphics[width=\linewidth]{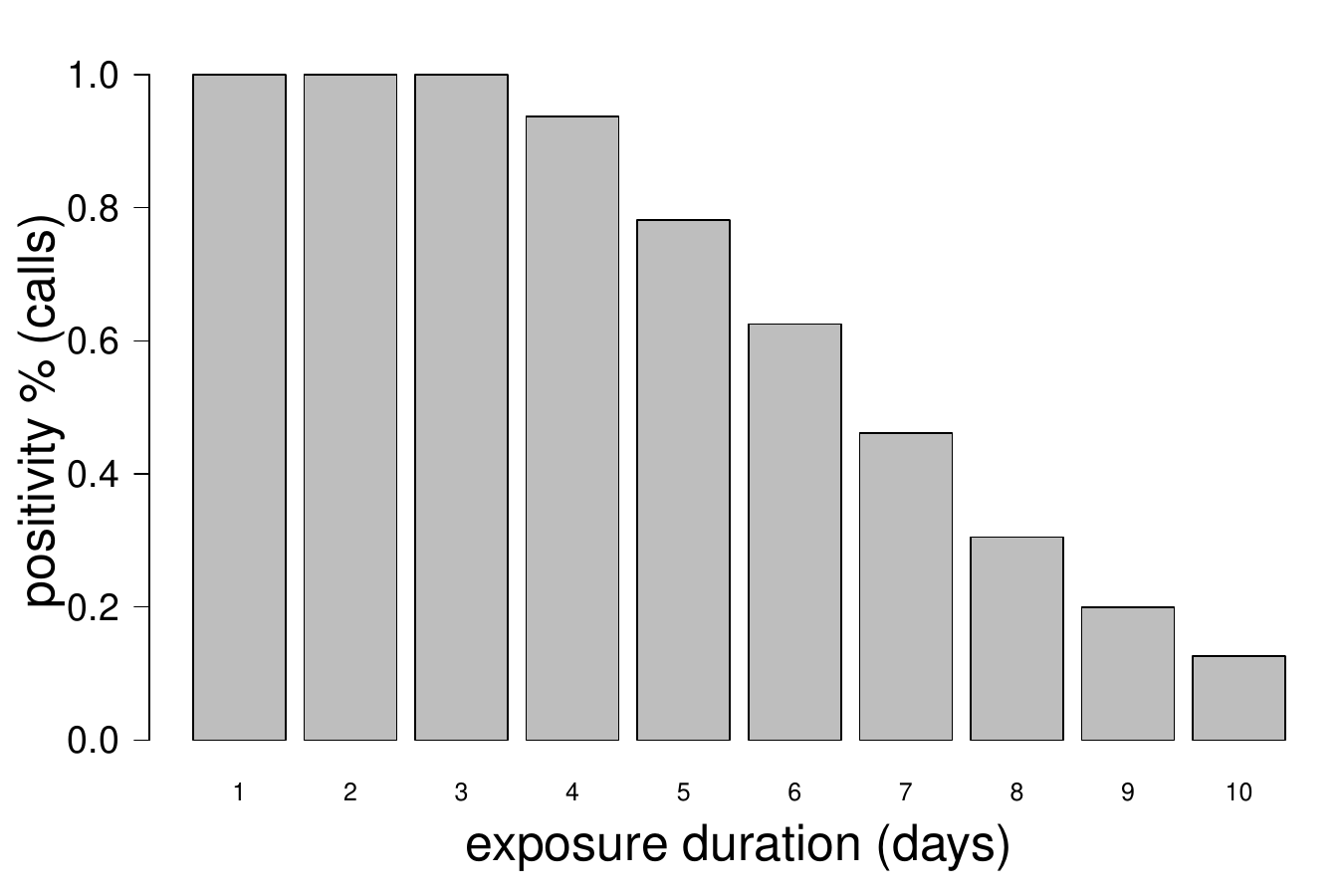}
\end{subfigure}\hspace*{\fill}
\begin{subfigure}{0.48\textwidth}
\includegraphics[width=\linewidth]{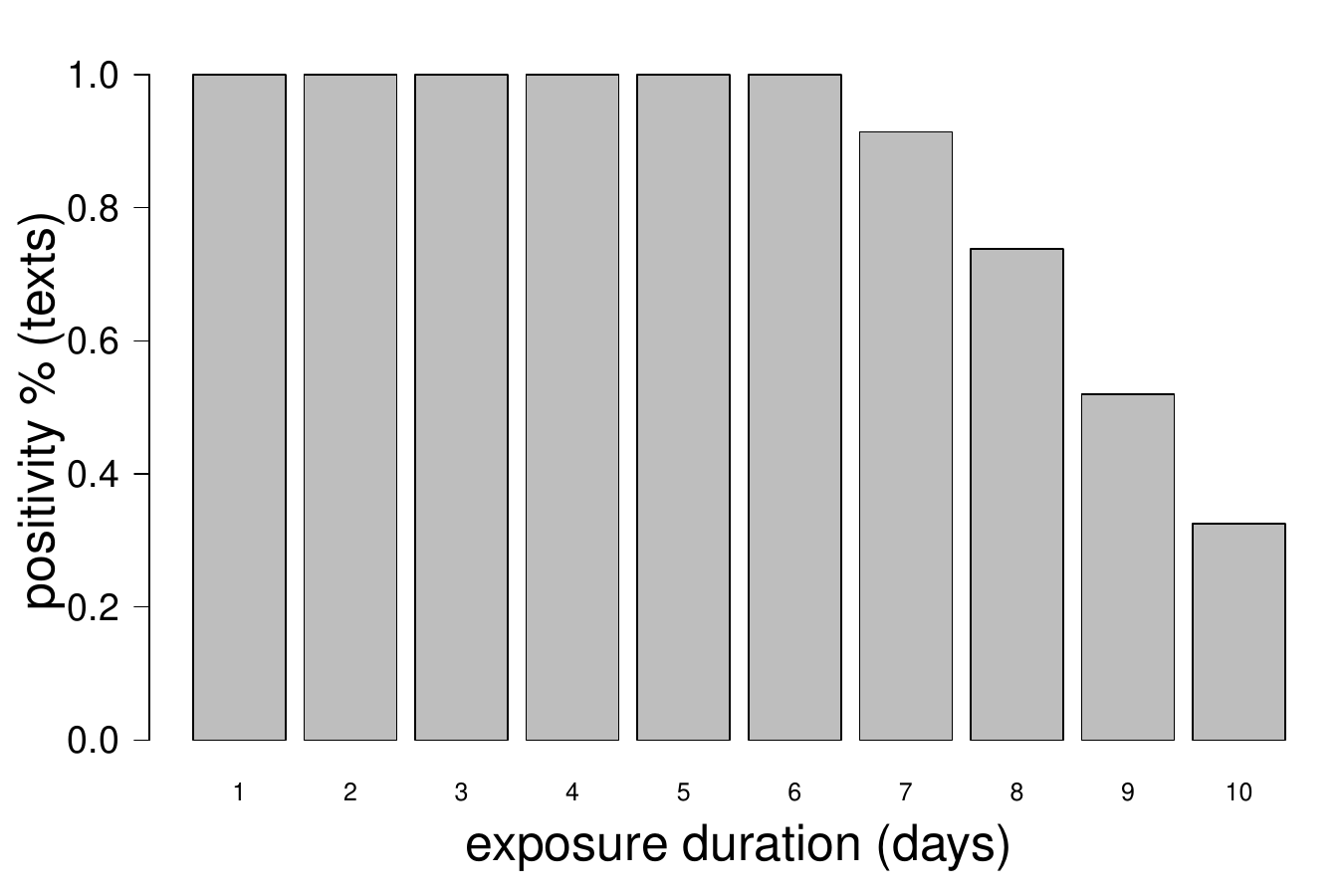}
\end{subfigure}
\caption{Positivity validation plots for phone-based social connectivity via calls (left) and texts (right) across a range of intervention durations.} 
\label{fig:positivity_validation}
\end{figure}

Figure~\ref{fig:positivity_validation} depicts the positivity validation plot for text and call connectivity across a range of interventional durations.
For call connectivity, all possible values are observed for exposures up to 3 days, while for text connectivity, this holds up to 6 days.
If we consider designing a text connectivity intervention strategy for 7 days under the constraints of no more than 3 times, 34 out of 35  unique intervention possibilities are observed (including the three strategies in Figure~\ref{fig:generaleffects}b except for $\ba_{(t-6):t}=(1,0,0,0,1,1,0)$. We can thus compare the estimated 7-step general effects for the 34 observed strategies, excluding the unobserved $\ba_{(t-6):t}=(1,0,0,0,1,1,0)$. 
R code for implementing this approach and associated diagnostic tools is available at the corresponding author's GitHub page.

\section{Discussion}
The rise in popularity of intensive longitudinal studies and multivariate time series data can be attributed to advancements in modern technology, which enables continuous, real-time monitoring via extensive measurements of subjects. 
The integration of mobile devices into medical, psychological, psychiatric, and social science research is a prime example of closely monitoring signals and biomarkers from individuals in their natural environment. 
Obtaining data-driven and well-informed policy decisions from this intricate data requires a comprehensive approach to causal inference.

This paper contributes to the growing methodological literature on causal inference for intensive longitudinal data and non-stationary multivariate time series. 
We propose a set of causal estimands designed for N-of-1 studies with time-varying effects in multivariate time series mobile device data. 
The causal estimands quantify the total and controlled direct effects of time-varying exposures on the outcome in both the short and long term as well as patterns of change over time. 
The identification strategy leverages the counterfactual approach to causal inference and the g-formula along with modeling assumptions within the state space modeling framework, which accommodate the non-stationary of the bio-behavioral processes and feedbacks between outcomes, exposures, and covariates. 
We assume neither static treatment effects nor stationary time series, and the identification strategy can be applied to both randomized trials and observational studies where the treatment assignments are confounded. 
The assumption of no unmeasured confounding is untestable and a key challenge in the application and interpretation of our approach. 
Further work is needed to develop sensitivity analysis strategies to address unmeasured confounding. 
Specifically, it would be of interest to leverage the negative controls that are naturally available in time series data to detect violations of the no unmeasured confounding assumption \cite{lipsitch2010negative,hu2023using}.
Our innovative approach fills a crucial gap in the literature, as it enables researchers to better capture dynamic treatment effects in personalized mobile health data. 

We introduce graphical tools, specifically the impulse impact plot and step response plot, to facilitate the illustration of the long-term impact of exposures over time. These approaches enhance the understanding of the temporal dynamics of treatment effects, allowing for clearer and more insightful interpretations of results. Ensuring the positivity assumption in causal inference is particularly challenging in the context of observational time series data. Traditional approaches ensure positivity through randomization or studying only a handful of time points. We propose a new diagnostic tool -- the positivity validation plot -- that evaluate the spectrum of treatment regimes that might be identifiable, representing valuable contribution to the field of causal inference in time series data.

The BLS is a pioneer study in long-term smartphone monitoring of multiple bio-behavioral processes in the context of Severe Mental Illness (SMI). The application of the proposed methodology sheds light on how phone-based social interaction may impact negative mood in both the short and long term. 
This analysis reveals the substantial heterogeneity in treatment effects over extended length of follow-up and across individuals in an N-of-1 framework, providing important implications for personalized interventions and treatment strategies in mental health care. 
Combining the positivity assumption validation plot and the ability of adjusting for patients' past treatment and health status, we also apply a simulation-based approach to infer personalized treatment recommendations. 
We underscore the practicability and utility of our approach in unraveling the causal relationships embedded within intricate N-of-1 time series data. Ultimately, our work offers valuable insights into shaping personalized treatment strategies in the realm of complex mobile device data.

In addition to violations of no unmeasured confounding assumption and the positivity assumption mentioned above, we acknowledge additional limitations in the of causal interpretation of our analyses. 
Specifically, our approach relies on correct model specification, including parametric assumptions and linear relationships. 
To reduce dependence on the model specification, developing matching approaches or propensity score based estimators for intensive longitudinal data is an important research direction. 

We also must note that psychiatric, psychological, and behavioral biomarkers are difficult to access and reliably quantify in natural settings \cite{Valeri2023digitalpsychiatry}, and are prone to a considerable amount of missing data. 
While mobile devices provide an innovative alternative option to overcome this long-standing barrier of reliable biomarker measurement in patients' daily lives through the passive collections of behavioral and environmental signals and active collection of psychological biomarkers via personal mobile devices. 
Furthermore, in a time of increasing interest in personalized medicine, an individual's response to a medication or behavioral intervention may not be well represented by a population means from traditional randomized trials, and N-of-1 design provides novel theoretical framework to estimate the individual treatment effect over time. 
These studies are still subject to missingness, including potential missing not at random mechanisms.
In other work we have proposed an approach for missing data imputation that can be applied in conjunction with the here proposed methodologies \cite{xiaoxuan2022SSMimpute}.

Finally, while the focus of this contribution is to provide methods for individual causal effects estimation for precision medicine, there is also interest in combining evidence of treatment effect across multiple individuals for more generalizable and robust inferences in future work.

\singlespacing
\bibliographystyle{unsrtnat}

\bibliography{Causal_inference_for_multivariate_time_series_in_N-of-1}

\end{document}